\documentclass[11pt]{article}
\usepackage{textcomp}
\usepackage{multirow}
\usepackage{enumerate}
\usepackage[a4paper,left=3cm,right=3cm,top=3cm,bottom=2.5cm]{geometry}

\usepackage{graphicx}
\usepackage[hyperfootnotes=false, linktocpage=true, colorlinks, citecolor=blue, linkcolor=blue, urlcolor=blue, filecolor=blue]{hyperref}
\usepackage[subrefformat=parens,labelformat=parens]{subfig}

\usepackage{amsfonts,amsmath}
\usepackage{multirow,multicol}
\usepackage{bbm}
\usepackage{xspace}
\usepackage{slashed}
\usepackage{parskip}
\newcommand{\CY}{\text{CY}}

\newcommand{\bZ}{\mathbbm{Z}}

\newcommand{\bP}{\mathbbm{P}}
\newcommand{\cO}{\mathcal{O}}

\usepackage{color}

\usepackage{hyperref}  
\usepackage{graphicx}  

\usepackage{exscale}  
\usepackage{amssymb}  
\usepackage{amsmath}  
\usepackage{amsthm}   
\usepackage{stmaryrd} 
\usepackage{mathrsfs} 
\usepackage{commath}  
\usepackage{dsfont}   
\usepackage{float}    
\usepackage[italicdiff]{physics}  
\usepackage{tensor}  
\usepackage{scalerel,stackengine}  

\usepackage{listings}

\usepackage[T1]{fontenc}
\usepackage{sourcecodepro}

\providecommand{\Comp}{\ensuremath{{\mathbb{C}}}}

\providecommand{\CP}[1]{\ensuremath{{{\mathbbm{P}}^{#1}}}}

\makeatletter
\newcommand\footnoteref[1]{\protected@xdef\@thefnmark{\ref{#1}}\@footnotemark}
\makeatother

\DeclareMathAlphabet\mathbcal{OMS}{cmsy}{b}{n}

\usepackage{xcolor}

\definecolor{codepurple}{rgb}{0.651, 0.0149, 0.643}
\definecolor{codegray}{rgb}{0.3,0.3,0.3}
\definecolor{numbergray}{rgb}{0.5,0.5,0.5}
\definecolor{codeblue}{rgb}{0.251, 0.471, 0.949}

\usepackage{cite}
\usepackage{hyperref}

\numberwithin{equation}{section}

\begin{document}

\begin{flushright}
CERN-TH-2020-205\\
LMU-ASC 48/20\\
 \end{flushright}
\begin{center}
\vspace{2cm}
{\Large\bfseries Moduli-dependent Calabi-Yau and \textit{SU}(3)-structure metrics\\[3mm] from Machine Learning}\\[5mm]
\vspace{1cm}

{\bf Lara B. Anderson}${}^{\,a,}$\footnote{lara.anderson@vt.edu}, {\bf Mathis Gerdes}${}^{\,b,}$\footnote{mathisgerdes@gmail.com}, {\bf James Gray}${}^{\,a,}$\footnote{jamesgray@vt.edu}, {\bf Sven~Krippendorf}${}^{\,b,}$\footnote{sven.krippendorf@physik.uni-muenchen.de}, {\bf Nikhil Raghuram}${}^{\,a,}$\footnote{nikhilr@vt.edu}, {\bf Fabian Ruehle}${}^{\,c,d,}$\footnote{fabian.ruehle@cern.ch}\\
{\small
\vspace*{.5cm}
${}^{a\;}$Department of Physics, Robeson Hall, Virginia Tech\\
Blacksburg, VA 24061, USA\\[3mm]
${}^{b\;}$Arnold Sommerfeld Center for Theoretical Physics, Ludwig-Maximilians-Universit\"at\\
Theresienstr.~37, 80333 Munich, Germany\\[3mm]
${}^{c\;}$CERN, Theoretical Physics Department\\
1 Esplanade des Particules, Geneva 23, CH-1211, Switzerland\\[3mm]
${}^{d\;}$Rudolf Peierls Centre for Theoretical Physics, University of Oxford\\
Parks Road, Oxford OX1 3PU, UK
}
\end{center}
\vspace{1cm}
\begin{abstract}
\noindent We use machine learning to approximate Calabi-Yau and $SU(3)$-structure metrics, including for the first time complex structure moduli dependence. Our new methods furthermore improve existing numerical approximations in terms of accuracy and speed. Knowing these metrics has numerous applications, ranging from computations of crucial aspects of the effective field theory of string compactifications such as the canonical normalizations for Yukawa couplings, and the massive string spectrum which plays a crucial role in swampland conjectures, to mirror symmetry and the SYZ conjecture. In the case of $SU(3)$ structure, our machine learning approach allows us to engineer metrics with certain torsion properties. Our methods are demonstrated for Calabi-Yau and $SU(3)$-structure manifolds based on a one-parameter family of quintic hypersurfaces in $\bP^4.$
\end{abstract}

\newpage

\tableofcontents
\setcounter{footnote}{0}

\section{Introduction}
\label{sec:Introduction}

Finding numerical approximations to metrics for the compact dimensions of string theory is a subject which by now has a long history within the literature. Knowledge of such metrics is not always necessary. In the case of Calabi-Yau (CY) compactifications, Yau's theorem~\cite{yau} and techniques from algebraic geometry allow many quantities of interest to be computed without explicit knowledge of the Ricci-flat metric on the extra dimensions~\cite{Green:2012pqa}. For example, the massless spectrum of fields and superpotential Yukawa couplings can be computed in a purely quasi-topological manner.

However, there are several quantities in the effective field theory for which knowledge of the metric is apparently indispensable. The kinetic terms of matter fields, for example, are determined by the K\"ahler potential. This is a non-holomorphic function that is therefore inaccessible with the methods of algebraic geometry. One needs explicit knowledge of the metric (and indeed often other structures) in order to compute this crucial aspect of the low-energy effective field theory.

Unfortunately, Ricci-flat metrics on CY three-folds, being high-dimensional structures with no continuous isometries, are seemingly prohibitively hard to find analytically. This has led to the development of a number of numerical, and other, approaches to computing these and related quantities in the literature~\cite{Headrick:2005ch,Donaldson:2005mat,Douglas:2006hz,Douglas:2006rr,Braun:2007sn,Braun:2008jp,Headrick:2009jz,Anderson:2010ke,Anderson:2011ed,Ashmore:2019wzb,Cui:2019uhy,Kachru:2018van,Kachru:2020tat,Tripathy:2020mjj}. One common feature which is seen in such work is that each computation of a metric is performed at one point in moduli space at a time. In many potential applications of such work, this is a serious limitation. For example, whenever the dynamics of fields, and hence the moduli dependence of the metrics, are relevant such a restriction is a serious impediment to progress. 

In particular, knowing how the metric changes with the moduli is important for understanding moduli dependent masses in relation to the swampland distance conjecture~\cite{Ooguri:2006in} and imprints of moduli in solutions to the electroweak hierarchy problem (e.g.~\cite{Blumenhagen:2009gk}).

The need for a numerical approach to obtaining the metric on compactification manifolds becomes even more acute in the more general arena of compactifications of $SU(3)$ structure. The methods of algebraic geometry are largely not available in these cases, and thus even quantities that can be addressed quasi-topologically in the CY case may require explicit knowledge of the compactification metric in this more general setting.
 In addition, the class of compactification-suitable $SU(3)$ structure solutions discovered thus far seem to generically suffer from the presence of small cycles in the geometry, violating the consistency conditions for a good supergravity approximation to the compactification. It is perhaps surprising, therefore, that there is essentially no work on numerical approximations to $SU(3)$-structure metrics in the string compactification literature.
 
Machine learning (ML), in particular through the recent advances in deep learning, offers a flexible approach to finding solutions of differential equations. In this paper, we will look at the application of machine learning techniques to both Ricci-flat metrics on CY manifolds (see~\cite{Ashmore:2019wzb} for some related work and~\cite{Ruehle:2020jrk} for a review on data science applications in string theory) and numerical approximations to metrics associated to more general $SU(3)$ structures.

We will explore a number of different but related approaches to learning moduli-dependent metrics with neural network (NN) approximations:
\begin{itemize}
\item Learning the CY metric from a K\"ahler potential which can be trained in either a supervised or unsupervised fashion (see Section~\ref{sec:kaehler}).
\item Learning the CY metric directly (see Section~\ref{sec:learningthemetric}).
\item Learning $SU(3)$ structure metrics (see Section~\ref{sec:NN-torsion}).
\end{itemize}

In more detail, we first construct NNs that interpolate between numerical approximations to Ricci-flat CY metrics computed at different fixed points in complex structure moduli space using Donaldson's algorithm~\cite{Donaldson:2005mat}. We then train metrics in an unsupervised fashion by optimizing loss functions measuring the deviation from Ricci flatness (i.e.~without using Donaldson's algorithm). For both types, the metrics are obtained from an ``algebraic K\"ahler potential''~\cite{Tian:1990aaa} and are learned as a function of moduli. We then discuss a method for constructing neural networks which output approximations to Ricci-flat metrics using ML techniques directly. We compare the efficiency of the methods we present with existing techniques for computing numerical approximations to CY metrics. We find improved performance in efficiency (i.e.\ given accuracy per computation time) in comparison to existing algorithms. 

Having studied the case of metrics of $SU(3)$ holonomy, we will then turn to apply machine learning techniques to the metrics associated to $SU(3)$-structures with non-trivial torsion. We will again present two approaches: one based around the use of an ansatz and the other concerning learning the metric directly. We verify these methods by reproducing the exact analytic results for an $SU(3)$-structure metric obtained by one of the authors in~\cite{Larfors:2018nce}. One of the reasons that the second of these approaches in particular appears to be promising is that, in choosing contributions to the loss function, we will describe how one can choose the non-zero torsion classes of the target $SU(3)$-structure. This contrasts with the analytic approaches that have appeared in the literature, which propose an ansatz for the forms defining an $SU(3)$-structure and then see which torsion classes that ansatz gives rise to. In the context of string compactifications, this leads to a shooting problem. Specific constraints on the torsion classes are imposed by the equations of motion of the theory, and there is no guarantee that any given proposed ansatz will turn out to give an $SU(3)$-structure with the required properties. Clearly, the approach that we will detail here avoids this issue. 

The rest of this paper is organized as follows.
In Section~\ref{sec:CYMetrics}, we discuss our setup to approximate the complex structure moduli dependence of Ricci-flat CY metrics. Section~\ref{sec:NN-torsion} describes how approaches similar to the previous sections can be used to learn $SU(3)$-structure metrics. We conclude and present an outlook in Section~\ref{sec:Conclusions}. Technical details about our experiments and implementation of known algorithms can be found in the Appendices.

As this work was being completed, we became aware of related work in~\cite{Douglas:2020hpv} which is coordinated to appear simultaneously.
\section{Ricci-flat CY metrics}
\label{sec:CYMetrics}

The uniqueness and existence of metrics with vanishing Ricci curvature on compact K\"ahler manifolds with vanishing first Chern class is long known.\footnote{Calabi conjectured that for a compact K\"ahler manifold with vanishing first Chern class there exists a K\"ahler metric in the same cohomology class with vanishing Ricci curvature~\cite{Calabi}. He proved uniqueness of this metric and, subsequently, Yau proved the existence of such a metric~\cite{yau}.}
In this section, we first highlight several important known results relevant for obtaining numerical approximations to these K\"ahler metrics, with the goal of setting up our notation and to introducing examples used in our ML approaches. We then discuss several approaches to finding numerical metrics using deep learning and compare their performance with existing techniques.

\subsection{Ricci flatness from a Monge-Amp\`ere equation}
The Ricci curvature on K\"ahler manifolds can be written in a simple form
\begin{equation}
 R_{i\bar{\jmath}}=-\partial_i\overline{\partial}_{\bar{\jmath}} \log{\det g}~,
\end{equation}
where the metric $g=\partial\overline{\partial}K$ is obtained from the K\"ahler potential $K.$ Solving this equation for a K\"ahler potential corresponds to solving a fourth order partial differential equation (PDE). The following idea reduces the problem to solving a second order Monge-Amp\`ere (MA) equation.

One starts with any K\"ahler metric $g$ on a CY $d$-fold and its associated K\"ahler form $J_g$. The Ricci flat metric $g_\CY$ with K\"ahler form $J$ is supposed to be in the same cohomology class. Hence, it can be written as
\begin{align}
\label{eq:MAEquationRaw}
J = J_g + \partial\overline{\partial} \phi
\end{align}
for some smooth zero-form $\phi$. The second order Monge-Amp\`ere equation arises by noting that there are two ways of building a (top) volume form. The first (top) volume form $J^3$ arises from the K\"ahler form $J$ corresponding to the Rici-flat metric. Another volume form is given in terms of the holomorphic $(3,0)$ form $\Omega$ on the CY manifold by forming $\Omega\wedge\bar\Omega$. Since the top form is unique, these need to be proportional:
\begin{align}
\label{eq:MAEquation}
J\wedge J\wedge J=\kappa\; \Omega\wedge\bar\Omega\,
\end{align}
for some $\kappa\in\mathbbm{C}$ that is constant at any given point in moduli space. Using~\eqref{eq:MAEquationRaw}, this becomes a Monge-Amp\`ere equation for $\phi$.

The deviation from this proportionality measures how close a given metric is to Ricci flatness. We will return to this later in this section, when we discuss accuracy measures in detail.

\subsection{CY example: quintic hypersurfaces}
\label{sec:cyexample}
While there are many CY spaces, and string theory applications desire techniques available for all of these spaces---in particular examples with larger Hodge numbers---we restrict ourselves to prototype complex $d$-dimensional hypersurfaces in $\bP^{d+1}.$ 
Specifically, we focus on the one-parameter family of hypersurfaces
\begin{align}
\label{eq:CYHypersurfaces}
p_\psi(\vec{z})=\sum_{i=0}^{d+1} z_i^{d+2} +\psi \prod_{i=0}^{d+1} z_i = 0\,,
\end{align}
where the parameter $\psi\in\mathbbm{C}$ encodes the complex structure dependence and where we denote the homogeneous coordinates
\begin{align}
\vec{z}=[z_0:z_1:\ldots:z_{d+1}]\,.
\label{eq:homogeneouscoordinates}
\end{align}

In line with the degree of this equation, CY one-folds (i.e.\ tori) of this type are called cubics, CY two-folds (i.e.\ K3 manifolds) of this type are called quartics, and CY three-folds are called quintics. Most of our subsequent discussion is focused on the one-parameter quintic hypersurfaces.

The holomorphic $(d,0)$ form $\Omega$ can be constructed straightforwardly for hypersurfaces or complete intersections in projective ambient spaces~\cite{Candelas:1987kf}. If we restrict to a patch where $z_a=1$ (i.e.\ pick a set of local affine coordinates) and consider the coordinate $z_{b}$ as an (implicit) function of the coordinates $z_{c}$ with $c\neq a,b$, the form $\Omega$ is given by
\begin{align}
\label{eq:Omega}
\Omega = \frac{1}{\partial p_\psi(\vec{z})/\partial z_{b}} ~\bigwedge_{\substack{c=1,\ldots,d\\ c\neq a,b}} dz_c\,,
\end{align}
where $p_\psi$ is the hypersurface constraint as introduced in~\eqref{eq:CYHypersurfaces}. 

In practice, we will use two types of conventions for coordinates on the CY manifold. The first choice is to stick to the full set of homogeneous coordinates, and pick the coordinate for which $|p_\psi(\vec{z})/\partial z_{b}|$ is largest as the one the defining equation is solved for. Alternatively, we can use affine coordinates in each different patch and pick the induced coordinate as before. For numerical stability, it is best to go to the affine patch where we scale the homogeneous coordinate with the largest absolute value to one, $|z_a|=1$, and pick the dependent coordinate as above. This convention defines the relation between the ambient space and local coordinates on the CY manifold by uniquely specifying the patch and the variables the CY hypersurface equations are implicitly solved for. Note that this choice of affine coordinates leads to values in a unit ball, which readily normalizes the input for our neural networks.

For sampling points on the CY space we use the method outlined in~\cite{Douglas:2006rr}. We fix a random line in $\bP^4$ by choosing two random points (with flat prior) on the unit sphere isomorphic to $\bP^4$. Intersecting this line with the hypersurface $p_\psi(\vec{z})=0$ gives five points on the CY manifold which we use as our sample. It should be noted that these points are not sampled with a flat prior on the CY manifold, which means that the points have to be weighted accordingly in the numerical Monte Carlo integration of some function $f$:
\begin{equation}
 \int_{X_\psi} f~d{\rm Vol}_{\rm CY}=\int_{X_\psi} f~\frac{d{\rm Vol}_{\rm CY}}{dA}dA\approx \frac{1}{N}\sum_{i=1}^N f(\vec{z}_i)w(\vec{z}_i)~,
\end{equation}
where each of the $N$ points is weighted with $w(\vec{z}_i)=\frac{d{\rm Vol}_{\rm CY}}{dA}|_{\vec{z}_i}.$ The numerator is evaluated with the value of the top form $d{\rm Vol}_{\rm CY}\propto\Omega\wedge\bar{\Omega}$ and the denominator can be obtained from the pullback of the Fubini-Study metric of the ambient space onto the hypersurface, $dA\propto i^*_{p} \omega^{\rm FS}_{\bP^4}$. We refer the reader to Appendix~\ref{app:Sampling} for more details.

The following discrete symmetries can reduce the number of independent components in the K\"ahler potential ansatz. The quintic hypersurfaces~\eqref{eq:CYHypersurfaces} enjoy a $\bZ_{d+2}\times\bZ_{d+2}$ freely acting symmetry
\begin{align}
\nonumber \bZ_{d+2}^{(1)}\;: [z_0:z_1:\ldots:z_{d+1}]&\mapsto [\alpha^0 z_0: \alpha^1 z_1:\ldots:\alpha^{d+1}  z_{d+1}]\,,\qquad \alpha=e^{2\pi i/(d+2)}\\
\bZ_{d+2}^{(2)}\;: [z_0:z_1:\ldots:z_{d+1}]&\mapsto [z_1: z_2:\ldots:z_{d+1}:z_0]\,,
\label{eq:discretesymmetries}
\end{align}
i.e.\ these symmetries act by multiplication of complex phases and by cyclic permutation. It should be noted that the full (non-free) symmetry group $(S_{d+2}\times\mathbbm{Z}_{d+2})\rtimes\mathbbm{Z}_{d+2}^d$ is much bigger (here, $S_{d+2}$ denotes the permutation group and $\rtimes$ a semi-direct product). We note that the most generic hypersurface will not have these symmetries and we hence do not enforce them in our ML models to keep our ansatz as general as possible.

\subsection{Metric ans\"atze}
\label{sec:ansatz}
There is a canonical choice of K\"ahler metrics in complex projective spaces called Fubini-Study (FS) metrics. For a given complex projective space $\bP^n$, the FS K\"ahler potential can be written as
\begin{align}
K=\frac{1}{2\pi}\ln(\mathbf{k})\,, \qquad \mathbf{k}=\sum_{a=0}^{n} z_a \bar{z}_a\,,
\end{align}
and the corresponding K\"ahler metric is
\begin{align}
g_{a\bar{b}}=\partial_a\overline{\partial}_{\bar{b}} K = \frac{1}{2\pi}~\frac{\mathbf{k} \mathbf{k}_{a\bar{b}}-\mathbf{k}_a \mathbf{k}_{\bar{b}}}{\mathbf{k}^2}\,,
\end{align}
where the subscripts on $\mathbf{k}$ denote differentiation. These FS metrics can be generalised by adding a Hermitian matrix $H$
\begin{equation}
 \mathbf{k}=\sum_{a,\bar{b}=0}^{n} z_a~H_{a\bar{b}}~\bar{z}_{\bar{b}}\,.
 \label{eq:generalisedFS}
\end{equation}
So-called algebraic metrics are obtained by considering non-trivial pull-backs of these generalised FS metrics~\eqref{eq:generalisedFS} defined in high-dimensional projective spaces to our ambient space (i.e.~$\bP^4$ in the case of the quintic). The embedding is constructed via global sections $s_\alpha$ of line bundles which are non-trivial on the CY:
\begin{equation}
\mathbf{k}=\sum_{\alpha,\bar{\beta}=0}^{N_k} s_\alpha(\vec{z})~H_{\alpha\bar{\beta}}~\bar{s}_{\bar{\beta}}(\vec{\bar{z}})\,.
\end{equation}
In practice, the basis of these sections is given by polynomials of degree $k$ and grows like $N_k\sim O(k^{d+1})$ for a $d$-dimensional CY manifold, i.e.~$N_k \sim O(k^4)$ for the quintic.\footnote{Polynomials containing the defining equations are removed. Details on the implementation are summarized in Appendix~\ref{app:Polynomials}.} 
An interesting aspect of this parametrization is that linear combinations of the global sections $s_\alpha$ at degree $k$ give the eigenfunctions corresponding to the first $k+1$ eigenvalues of the scalar Laplacian on $\bP^4$, cf.~\cite{Braun:2008jp}.\footnote{The eigenvalues of the scalar Laplacian on $\bP^4$ have increasing multiplicities due to the fact that they carry representations of the $SU(4)$ action on the projective space.}
In this sense, the algebraic metrics can be understood as spectral expansions with coefficients given by the $H$-matrix.

Donaldson's algorithm~\cite{Donaldson:2005mat} provides a method which determines $H$ for any given $k$ such that $H$ is balanced. In the limit $k\to\infty$, these balanced metrics are unique and converge to the Ricci-flat CY metric. Details about Donaldson's algorithm, more definitions, and our implementation can be found in Appendix~\ref{sec:Donaldson}.

\subsection{Accuracy measures}
\label{sec:Accuracy}
Measuring how close a given metric is to the Ricci-flat CY metric is useful for two reasons:
\begin{enumerate}
 \item One can check the convergence of the numerical method and compare different numerical approximations.
 \item One can optimize the metric by minimizing these measures. In different words, if one uses these measures as loss functions, finding Ricci-flat CY metrics is readily defined in the language of ML.
\end{enumerate}
In order to evaluate the quality of an approximation, the authors of~\cite{Douglas:2006rr} propose to compute the quantity
\begin{align}
\label{eq:eta}
\eta = \frac{J\wedge J\wedge J}{\Omega\wedge\bar\Omega}=\frac{(-6i)\det g}{\Omega\wedge\bar\Omega}\,,
\end{align}
which should be constant throughout the CY manifold. In fact, this should just be equal to $\kappa$ defined in Equation~\eqref{eq:MAEquation}. Note that $J$ here can be any K\"ahler form in the same K\"ahler class (e.g.~the Fubini-Study metric) and need not be the K\"ahler form of the Ricci-flat metric. As an accuracy measure, the authors propose to compute
\begin{align}
\label{eq:sigma}
\sigma=\frac{1}{\int_{X}\Omega\wedge\bar\Omega}~\int_X\left|1-\frac{1}{\kappa}\;\frac{J^3}{\Omega\wedge\bar\Omega}\right|\,.
\end{align}
Hence, if $\eta$ is constant at each point on the CY space, the integrand vanishes and $\sigma=0$. In practice, our losses can be of the generalized form 
\begin{equation}
 \mathcal{L}_{\rm MA}=\alpha \left|1- \frac{1}{\kappa} \frac{J^3}{\Omega\wedge\bar{\Omega}}\right|^n,
 \label{eq:algconstloss}
\end{equation}
which allows for some overall re-scaling of $J$ or $\Omega$, the option of doing a Monte-Carlo approximation of the $\sigma$ accuracy. Note that a larger $n$ punishes outliers in this measure more strongly. In Section~\ref{sec:kaehler}, we generally follow the authors of~\cite{Douglas:2006rr} and do not rescale $J$ or $\Omega$ to set $\kappa=1$. When learning the metric directly in Section~\ref{sec:learningthemetric}, we learn it in a normalization such that $\kappa=1$, i.e.\ we force the metric networks to learn the rescaled metrics since we keep $\Omega$ fixed.

Alternatively, one can use the vanishing of the Ricci scalar as an accuracy measure when we learn the K\"ahler potential directly. However, using two additional derivatives takes longer and the numerics appeared to be less stable and accurate.

In addition to being Ricci flat, the solution has to be K\"ahler. Of course, if we learn a K\"ahler potential, this property is guaranteed, so we only need to impose it when learning the metric directly. The condition is that the fundamental two-form is closed,
\begin{equation}
 dJ=0\quad\leftrightarrow \quad g_{i\bar{\jmath},k}~dz_i\wedge d\bar{z}_{\bar{\jmath}} \wedge dz_{k}=0=g_{i\bar{\jmath},\bar k}~dz_i\wedge d\bar{z}_{\bar{\jmath}} \wedge d\bar{z}_{\bar k}~.
\end{equation}
This leads to $9$ non-trivial complex or respectively $18$ real conditions
\begin{equation}
c_{ijk}=g_{i\bar{\jmath},k}-g_{k\bar{\jmath},i}=0~.
\end{equation}
We implement these conditions by taking derivatives of the NN with respect to the input variables. Note that this is different from the usual backpropagation in neural networks, where derivatives are taken with respect to the parameters of the neural network layers. As the induced coordinate, i.e.\ the coordinate which is implicitly specified in terms of the other coordinates upon imposing the hypersurface constraint, is an additional input to our network, we need to properly take this into account when taking derivatives. We have implemented each of these $18$ conditions $c_i$ for our networks. We measure the K\"ahlerity accuracy as follows
\begin{equation} \label{Kloss}
 \mathcal{L}_{\rm dJ}=\sum_{i,j,k} ||{\rm Re}(c_{ijk})||_n+ ||{\rm Im}(c_{ijk})||_n~,
\end{equation}
where in our experiments we have used both $n=1$ and $n=2.$ A good cross-check which we used to test our implementations is that this K\"ahler loss is zero for the FS metric.

The third consistency condition we need to impose is that the metric transforms correctly on overlaps of patches of the projective ambient space. The K\"ahler potential ansatz in Section~\ref{sec:ansatz} automatically satisfies the overlap conditions, so these conditions are primarily used when we go beyond the ansatz in Section~\ref{sec:learningthemetric}. Defining the standard patches $U_i=\{z_i\neq0\}$, we can use the projective scaling to set $z_i=1$ and obtain an affine patch with coordinates
\begin{align}
\vec{z}^{(i)}=\left(\frac{z_0}{z_i},\frac{z_1}{z_i},\ldots,\frac{z_{i-1}}{z_i},\frac{z_{i+1}}{z_i}\ldots,\frac{z_{d+1}}{z_i}\right)\,.
\label{eq:affinecoordinates}
\end{align}
The transition function from $U_i$ to $U_j$ is then simply $z_i/z_j$. This allows us to compute the transition functions for $g$. Denoting the transition matrix with $T_{ij}=\partial \vec{z}^{(i)}/\partial \vec{z}^{(j)}$, we can compute the metric $g^{(j)}$ in patch $j$ from the metric $g^{(i)}$ in patch $i$ via
\begin{align}
\label{eq:TransitionFunction}
g^{(j)}=T_{ij}\cdot g^{(i)}\cdot T_{ij}^\dagger\,.
\end{align}
Almost all points\footnote{That is, all points up to a measure zero set.} lie in all standard affine patches $U_i$ of $\bP^{d+1}$. Hence, if we use different patches as inputs to describe the same point on the CY manifold, the metric should transform as dictated by the transition functions between the patches.

In order to compute the overlap loss, we proceed as follows. As explained above, we usually go to the patch $U_i$ where $i$ is the index of the coordinate of a point on the CY manifold with the largest absolute value, and we solve for $z_j$, where $j$ is the index of the coordinate which has the largest absolute value of the derivative $\partial_j p_\psi$. Typically, in the input for the NN we already divide all coordinates by $z_i$. Now, we will also input the coordinates of the point in question in other patches $U_k$ with $k\neq j$ and compute the resulting expression for the metric based on this NN input. We then compute the expected value in the patches $U_k$ using the transition functions and write the transition loss as
\begin{align}
\label{transloss}
\mathcal{L}_\text{Transition}=\frac1d \sum_{k,j} \big|\big|g^{(k)}_{\rm NN}(\vec{z}) - T_{jk}(\vec{z})\cdot g^{(j)}_{\rm NN}(\vec{z})\cdot T_{jk}^\dagger(\vec{\bar{z}})\big|\big|_n\,,
\end{align}
where the transition functions are as explained in~\eqref{eq:TransitionFunction} and the numerical values of points we use are explained when we describe our metric networks in Section~\ref{sec:learningthemetric}.
 We define the matrix norm for $n>1$ via the sum of all matrix components,
\begin{align}
||M_{\mu\nu}||_n=\sum_{\mu,\nu} |M_{\mu\nu}|^n\,,
\label{eq:matrixnorm}
\end{align}
and for $n=1$ as the sum of the absolute values of all matrix elements $M_{\mu\nu}$. Finally, we sum in the loss over all patches (except for $j$). Since we compute the loss for $d$ overlaps, we introduce a conventional factor of $1/d$ into the loss. Again, this loss is non-negative and goes to zero if the metric transforms correctly across patches. In order to cross-check the code, we can again use the Fubini-Study metric, which is well-defined on the overlaps.

\subsection{Finding metrics with machine learning}
\begin{figure}[t]
  \centering
  \includegraphics[width=0.95\textwidth]{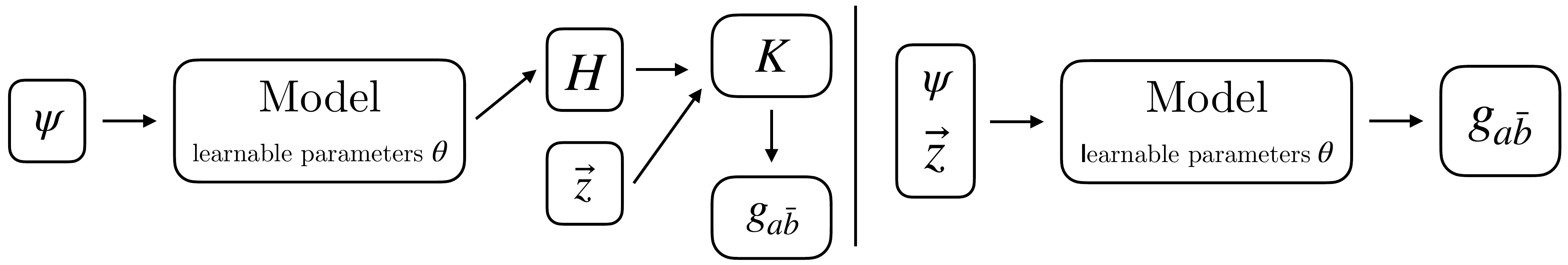}
  \caption{Schematic overview of how models predicting the Hermitian matrix  $H$ ({\bf left}) and the metric $g_{a\bar{b}}$ ({\bf right}) are designed. The respective models are neural networks of different complexity.}
  \label{fig:overview}
\end{figure}

Given the accuracy measures just discussed, it is clear that finding Ricci-flat metrics can be formulated as a continuous ML optimization problem of the underlying algebraic and differential equations. The first implementation choice one has to make is whether to learn representations of the K\"ahler potential or the metric. A schematic overview of either setup can be found in Figure~\ref{fig:overview}. The present ML approach is in large parts facilitated by readily available frameworks implementing auto-differentiation, as this allows one to optimize the appropriate loss functions which involve derivatives of the K\"ahler potential and metric respectively.\footnote{We utilize PyTorch~\cite{NEURIPS2019_9015}, Tensorflow~\cite{DBLP:journals/corr/AbadiABBCCCDDDG16}, and JAX~\cite{jax2018github} in our experiments.}

The motivation for an ML approach is that an improvement in speed and accuracy by using these numerical methods enables a study of CY metrics at a much broader scope. The current numerical benchmark is given by Donaldson's algorithm, which computes metrics at a fixed point in moduli space and becomes significantly more expensive when constructing more accurate metric approximations in the sense of Section~\ref{sec:Accuracy}.

A first approach is to use NNs for supervised regression on  K\"ahler potentials using the output of Donaldson's algorithm as discussed in Section~\ref{sec:supervised_donaldson}. Alternatively, instead of performing fixed point iteration of the T-operator in Donaldson's algorithm, we can use the same ansatz for the K\"ahler potential, but utilize the $\sigma$-accuracy measure to directly optimize the output of our neural network. While Donaldson's algorithm is guaranteed to converge for $k\to\infty$, for finite, fixed $k$ there exist better approximations (as quantified by the flatness measure $\sigma$) than the ones obtained from Donaldson's algorithm. We also demonstrate that using the more expensive Ricci scalar as a loss is feasible (cf.~Appendix~\ref{sec:ricciloss}). Although this is similar to the approach in~\cite{Headrick:2009jz}, our approach takes into account the moduli dependence of the $H$-matrix as an input to our neural network (cf.~Section~\ref{sec:predictingHdirectly}). We stress that altering the setup to include multiple complex structure moduli is straightforward in terms of the architecture. In principle, one can also start with a different ansatz for the K\"ahler potential in the neural network, which we do not pursue further in this article. Instead, we learn the metric directly which we discuss in Section~\ref{sec:learningthemetric}.

The advantage of learning the K\"ahler potential is that it automatically satisfies $dJ=0$ and in the case of the algebraic metric ansatz the overlap conditions are guaranteed. The advantage of learning the metric $g$ directly is that it is more general, for instance allowing for larger functional flexibility (e.g.~ability to capture solutions with $dJ\neq 0$). Moreover, learning the metric directly requires only learning the independent components of the hermitian $d\times d$ metric, while the ansatz for the K\"ahler potential requires dealing with matrices whose number of components $N_k^2$ grows rapidly.  

The feed-forward neural networks are implemented with standard packages. However, the loss functions associated to the accuracy measures are custom implementations. It is also worth noting that, when learning the metrics directly, we are not dealing with a supervised learning approach. Indeed, we do not know the CY metrics and hence cannot provide labels for supervised learning. Instead, the loss functions encode the continuous optimization task needed to solve the equations that ensure that the resulting metric is CY. In particular, we implemented the transition function computations as well as the matrix multiplication and the complex derivatives in terms of real and imaginary parts of the NN output $g$ and the inputs $z_i$ in order to be able to back-propagate in the optimization step through the respective losses. This splitting into real and imaginary parts is required in Tensorflow and PyTorch but can be avoided by using JAX.

\subsection{Learning the K\"ahler potential}
\label{sec:kaehler}
As mentioned above, learning the $H$-matrix as a parametrization of K\"ahler potentials has several advantages:
\begin{itemize}
\item The CY metric is guaranteed to be complex K\"ahler.
\item The CY metric is by construction globally defined, i.e.\ it glues nicely across different patches of $\bP^4.$
\item The resulting K\"ahler potential is given explicitly in terms of the sections $s_\alpha$, and consequently in terms of the coordinates $z_a.$
\end{itemize}
However, the disadvantage is that the $N_k\times N_k$ matrix $H$ has $N_k^2$ real independent entries, and $N_k$ grows rapidly with larger $k$ as can be seen from~\eqref{eq:numberSections}. Hence, this approach requires more and more training data in order to fix the coefficients and allow the interpolating NN to learn the complex structure dependence efficiently. It should be noted that  discrete symmetries can tremendously reduce the number of complex structure parameters. Moreover, equivariance of $s\cdot H \cdot \bar{s}$ can force many entries of $H$ to zero or to be equal. For example in the case of the quintic~\eqref{eq:CYHypersurfaces} with $k=2$, we find from~\eqref{eq:numberSections} that $H$ is a $15\times15$ matrix. Due to $H$ being Hermitian, this matrix has $15^2=225$ independent real components. However, the (multiplication by a complex phase and permutation) symmetries force the off-diagonal entries to be $0$ and impose relations between the diagonal entries. This leaves only two real degrees of freedom in $H$. It is a nice cross-check that the off-diagonal components automatically become zero and the respective diagonal components automatically have the same numerical values in Donaldson's algorithm, even if they have not been chosen to for the initial matrix $H^{(0)}$ (see Appendix~\ref{app:donaldsonelements}).

\subsubsection{Supervised regression of Donaldson's K\"ahler potentials}
\label{sec:supervised_donaldson}

First, we demonstrate that it is easily possible to train a NN to learn the moduli dependence in a supervised learning setup. This provides a simple check that the NN architecture has sufficient functional capabilities to learn the moduli dependence of $H$. To this end, we compute the matrix $H$ for different choices of complex structures using Donaldson's algorithm. The input to the NN is just the real part, imaginary part, and absolute value of $\psi$ and the output are the independent real and imaginary components of the Hermitian matrix $H$.

\begin{figure}[t]
\centering
\includegraphics[width=0.24\textwidth]{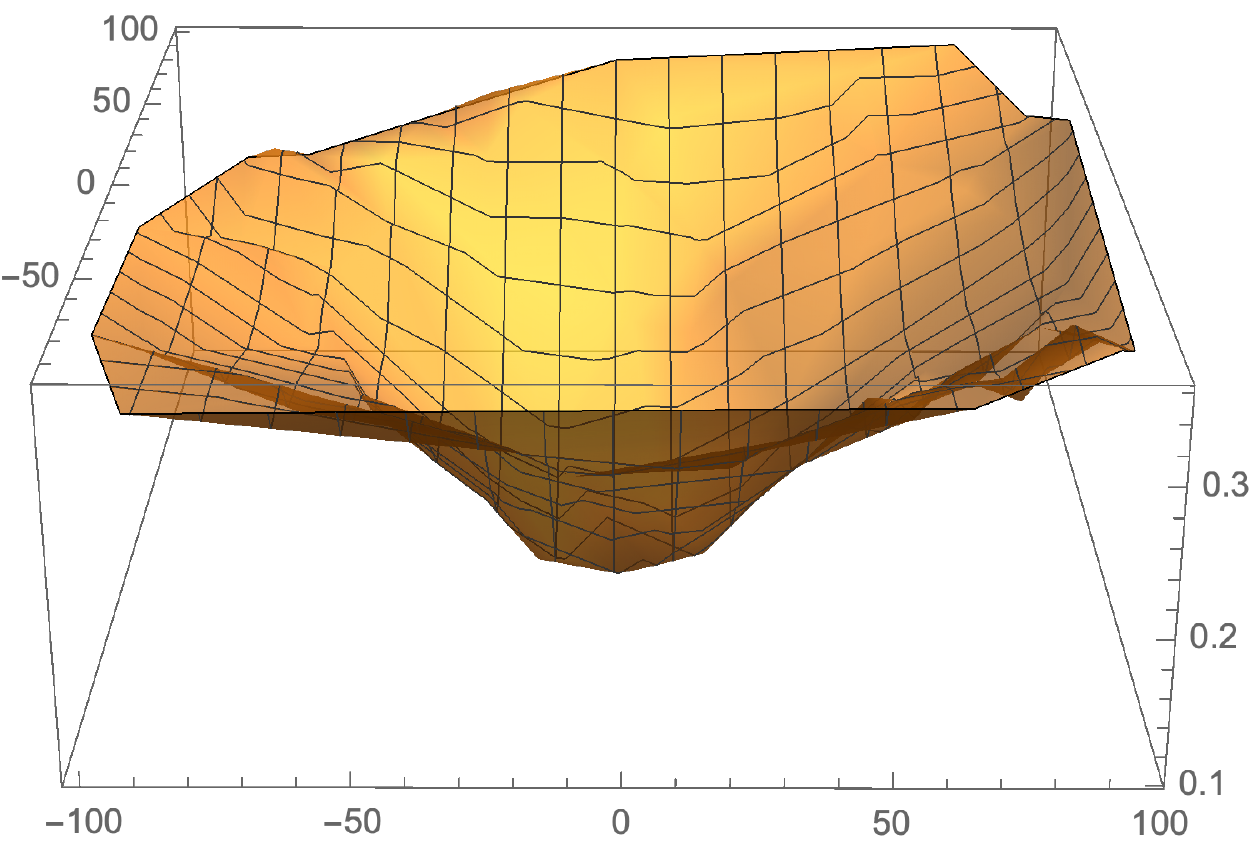}~
\includegraphics[width=0.36\textwidth]{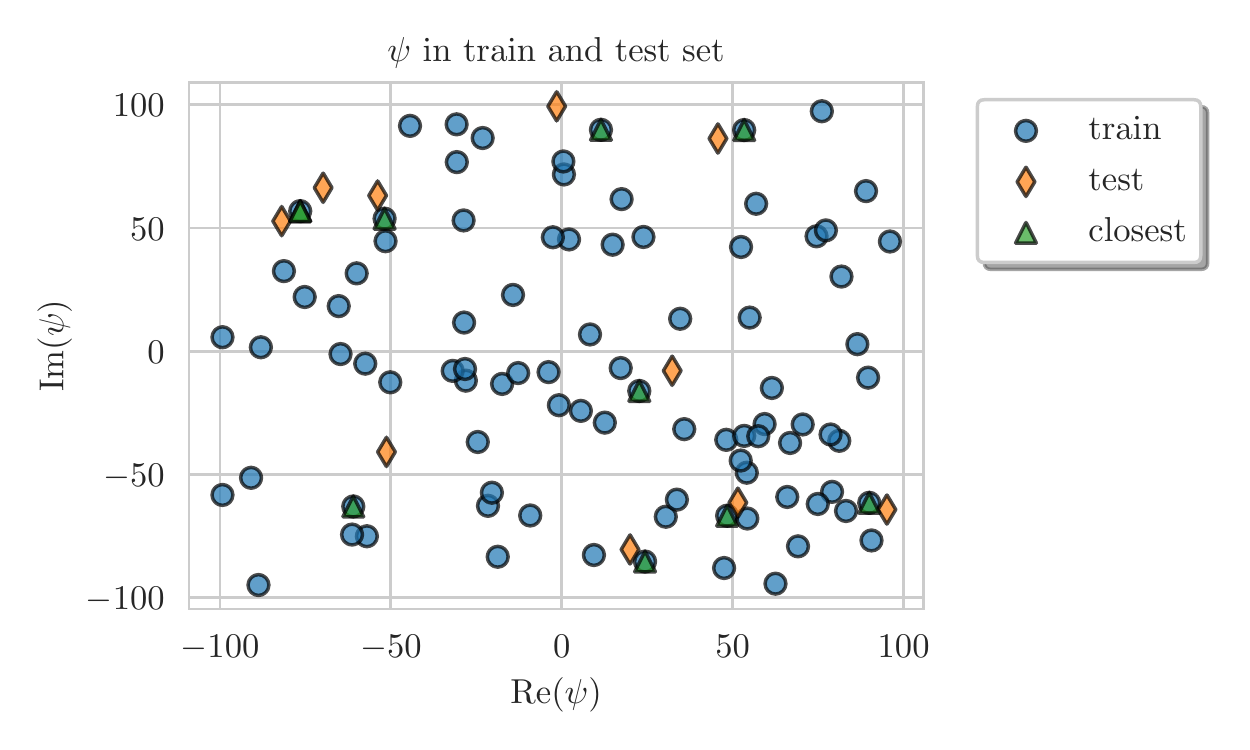}~
\includegraphics[width=0.36\textwidth]{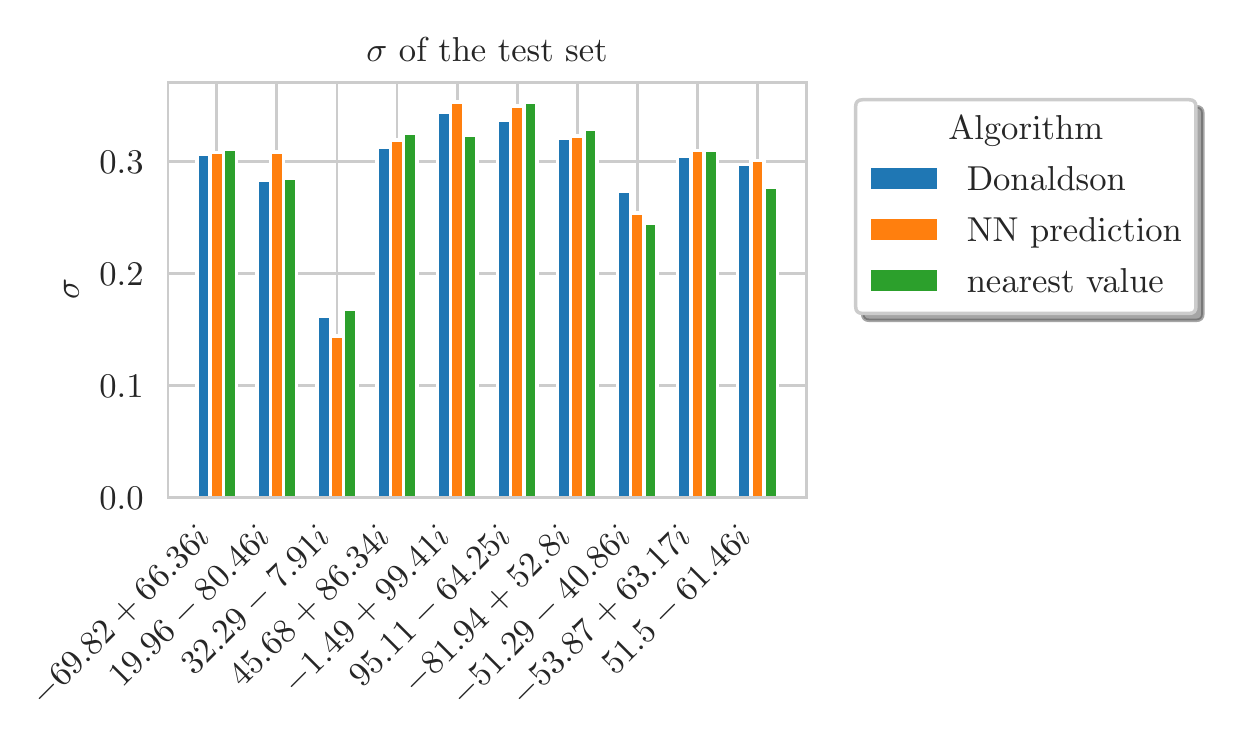}
\caption{(Left): Errors $\sigma$ for Donaldson's algorithm as a function of $\psi$ for random sampling. (Middle): Values of $\psi$ used in the training and evaluation set. (Right): Comparison of $\sigma$ for different approximation approaches.}
\label{fig:learnHrandom}
\end{figure}

{\bf Experiments}: We present results for the quintic~\eqref{eq:CYHypersurfaces} with $k=3$, which has $N_k=35$. We computed $H$ for different values of $\psi$ using Donaldson's algorithm using $80000$ points. In one experiment, we randomly drew 100 values for $\psi$ from a flat prior with $-100\leq\text{Re}(\psi)\leq100$ and $-100\leq\text{Im}(\psi)\leq100$ (see Figure~\ref{fig:learnHrandom} middle). We assess the quality of the NN interpolation by comparing the error measure $\sigma$ obtained by the NN on the test set with the result one would obtain by using the ``wrong'' K\"ahler potential computed for a point of the training set that is closest in complex structure moduli space (in Euclidean distance). For reference, we also compare these results with the result that is obtained from computing the K\"ahler potential at each point in the test set. In theory, this should provide a lower bound for the quality of the approximation that can be obtained from the NNs or from using a wrong but close-by approximation. In practice, as alluded to above, Donaldson's algorithm does not produce the K\"ahler potential with the lowest possible $\sigma$ error at fixed $k$, and we find that sometimes the NN and/or the nearest points produce better results than a direct computation following Donaldson's algorithm.

\begin{figure}[t]
\centering
\includegraphics[width=0.24\textwidth]{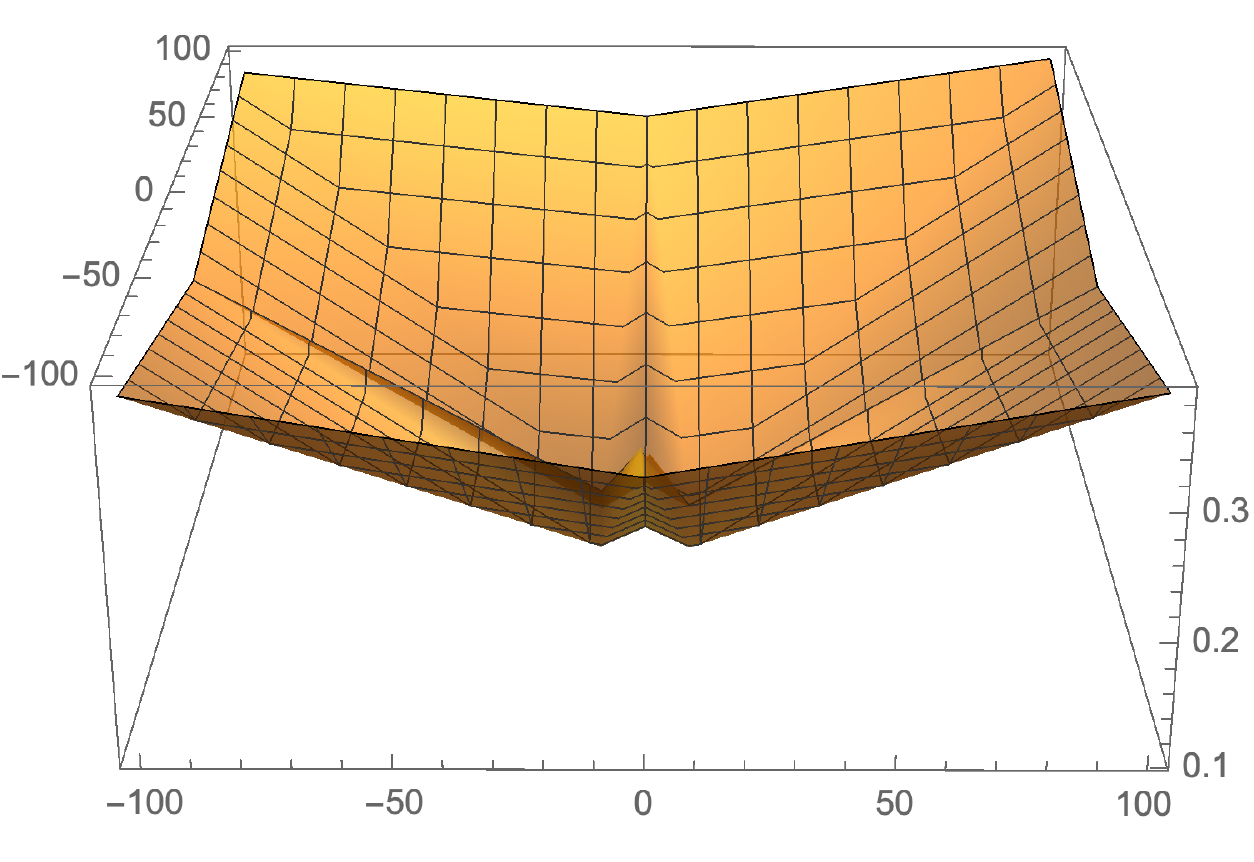}~
\includegraphics[width=0.36\textwidth]{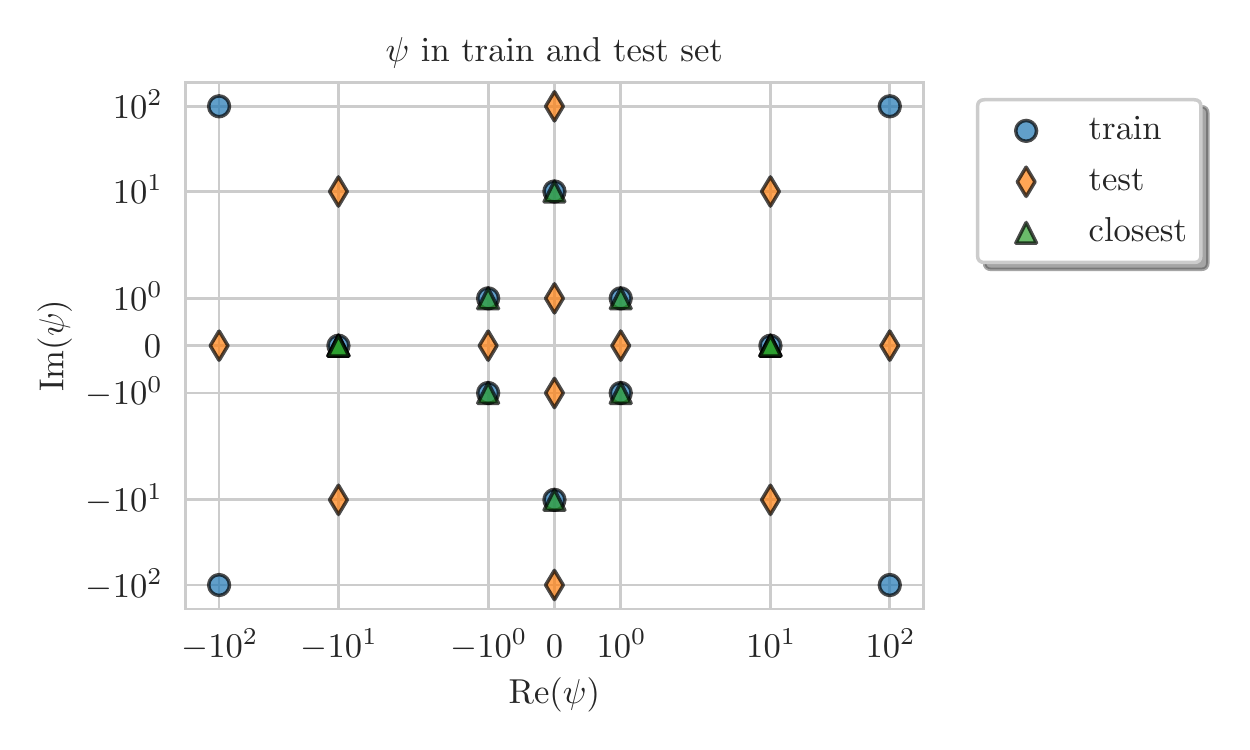}~
\includegraphics[width=0.36\textwidth]{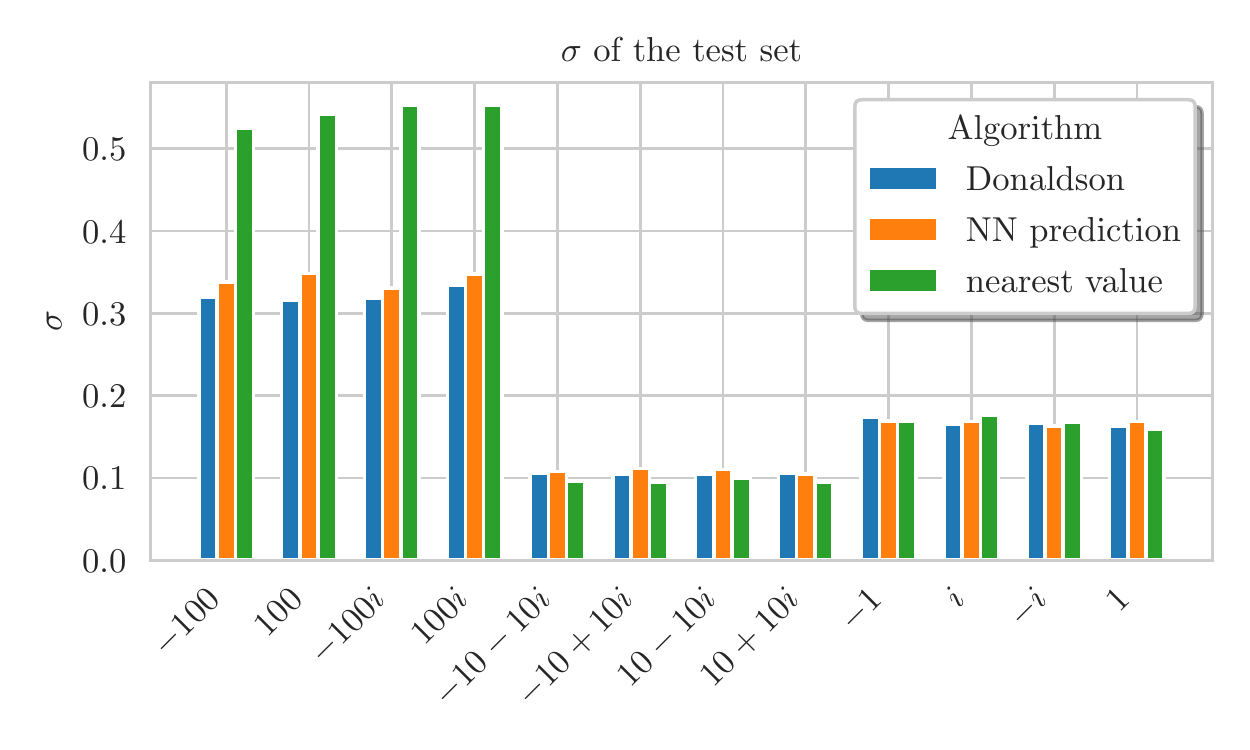}
\caption{(Left): Errors $\sigma$ for Donaldson's algorithm as a function of $\psi$ for sparse sampling. (Middle): Values of $\psi$ used in the training and evaluation set. (Right): Comparison of $\sigma$ for different approximation approaches.}
\label{fig:learnHbox}
\end{figure}

We also repeat this analysis, but this time we sample from a grid of complex structure values of the form $\psi=a+ib$ with $a,b\in\{0,\pm1,\pm10,\pm100\})$ (see Figure~\ref{fig:learnHbox} middle). Given that computing $H$ is very costly, especially for larger $k$, this grid contains only very few samples as compared to the rather dense sampling used in the first experiment.


We use feedforward neural networks with 3 hidden layers and ReLU activation. The input layer is three-dimensional, as it takes the real part, imaginary part, and the absolute value of $\psi$. The output layer is $N_k^2$-dimensional and gives the independent entries of $H$. We use ADAM as an optimizer~\cite{kingma2017adam}. Overall, we find that the results do not depend stron gly on hyperparameter choices such as the learning rate, the network architecture or activation function, the optimizer, the batch size, etc. Further details are given in Appendix~\ref{app:hdirectly_details}. 

The results of our two experiments are presented in Figures~\ref{fig:learnHrandom} and~\ref{fig:learnHbox}. On the left, we can see the error measure $\sigma$ in the complex $\psi$-plane. We find that $\sigma$ increases by a factor of 2 between $\psi\sim\cO(1)$ and $\psi\sim\cO(100)$. This illustrates that for larger complex structure, we need to go to larger $k$ in order to achieve the same quality of approximation of the Ricci-flat metric. However, interestingly, the error does not increase monotonically with $|\psi|$; indeed, for very small non-zero $\psi$ the error is larger than for $\psi$ in the intermediate range.\footnote{This happens throughout the complex $\psi$ plane, which is why we do not think that this is related to the conifold point.}  In the figure in the middle, we show the training (blue) and test (orange) sets for $\psi$. On the right, we plot $\sigma$ for all points in the evaluation set as obtained from Donaldson's algorithm, from using the $H$ as computed by the NN, and from using the wrong $H$ as computed via Donaldson's algorithm for the closest available $\psi$ in the training set.

As one can see for the randomly sampled points in Figure~\ref{fig:learnHrandom}, the differences for this rather fine sampling of points in the complex structure plane are not very large. While this means that the NN works well, it also shows that the error one makes by taking the $H$ that has been obtained from Donaldson's algorithm for a nearby point is not too large either. This changes if the sampling of points in the complex structure plane becomes more sparse, as shown in Figure~\ref{fig:learnHbox} (note the log scale on the axes). In that case, for the innermost ``square'' with $\{a,b\}\in\{\pm1,\pm1\}$, the results of using the nearest neighbor and the NN are small, as to be expected from our results for densely sampled, randomly distributed~$\psi$. However, as the distance between the nearest neighbors and the actual complex structure point increases, the nearest neighbor approximation becomes much worse than the NN prediction, as can be seen from the outer square. This illustrates that the NN can already learn a reasonable approximation of the functional form of the $\psi$ dependence of the coefficients in $H$ from a relatively small sample, which is important since computing $H$ is very time-consuming. Of course one could compare the NN against interpolation/regression algorithms other than neural networks. However, given the ease of implementing the NN, the extremely fast training time, and the quality of the results we refrained from exploring this further.

\subsubsection{Learning the Hermitian matrix $H$ directly}
\label{sec:predictingHdirectly}
In the previous section, we have shown that it is in principle feasible to train networks that approximate the Ricci-flat metric by learning the $H$ matrices produced by Donaldson's algorithm.
While we have seen that only few data are needed for the supervised training to produce useful interpolations, the approach is still limited by the accuracy and corresponding computational cost of Donaldson's algorithm.

Instead of building on top of Donaldson's algorithm, we now study networks that are trained directly using the Monge-Amp\`ere loss defined in Equation~\eqref{eq:algconstloss} with $n=2$. This is similar to the approach of \cite{Headrick:2009jz}, with the key difference that we are learning the moduli dependence of $H$. We find that this approach produces better accuracies while taking similar amounts of time as Donaldson's algorithm over a range of $\psi$ values.

We performed several experiments to find the best NN architecture to model the maps from $\psi$ to $H$. Based on several experiments, we have found the following architecture to work well. As input we take $|\psi|$ and the complex angle $\text{arg}(\psi)$. This is followed some number of dense layers using the sigmoid activation function. Finally, another dense layer (without activation) is added that maps to the needed number of complex parameters of $H$. We have found that the Cholesky decomposition typically leads to better results than encoding the real and imaginary parts directly, which ensures the output $H$ is positive definite in addition to Hermitian. Further technical details can be found in Appendix~\ref{app:hdirectly_details}. 

As a balance between numerical cost and quality of approximation, our experiments here are performed for $k=6$, which corresponds to $42025$ independent components in $H$ (as a general Hermitian matrix). We expect generally that optimal $H$ matrices exist for each degree $k$ and value of $\psi$ that converge to the Ricci-flat metric at a significantly faster rate than Donaldson's balanced metrics~\cite{Headrick:2009jz}. Our networks could in principle find these optimal values of $H$. However, their ability to do so will be limited by the complexity of the model that maps from $\psi$ to $H$, and by the range of $\psi$ values over which we optimize.
In an initial experiment where the network was optimized on uniform values $0<|\psi|<10$, the network reached a $\sigma$ accuracy comparable to Donaldson's algorithm at degree $k=12$ (see \figref{fig:hpsi-smallscale} in the appendix).
This is a noteworthy result, as training the network at $k=6$ over the whole range of $\psi$ takes only on the order of minutes, while Donaldson's algorithm at $k=12$ takes on the order of days using the same hardware.

\begin{figure}[t]
  \centering
  \includegraphics[width=\textwidth]{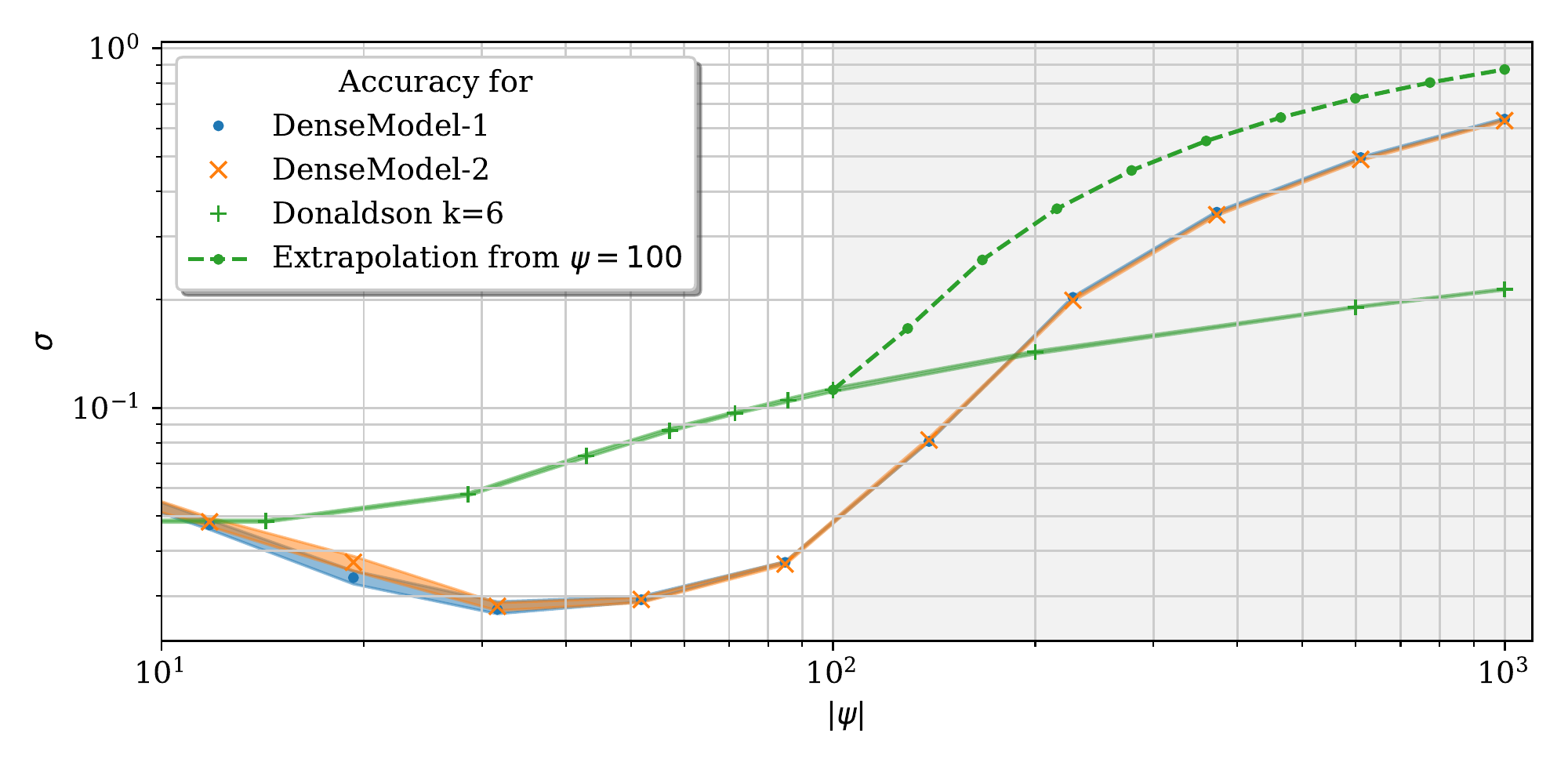}
  \vspace{-10pt}
  \caption{$\sigma$ accuracies at $k=6$ achieved by the dense network with one and two hidden layers. The shaded area indicates the range of $|\psi|$ that was not used during training, and thus shows the extrapolation behavior of the networks. For reference, the $\sigma$ accuracy achieved by Donaldson's algorithm for the same range of $|\psi|$ is shown. The dashed line corresponds to the extrapolation of using Donaldson's balanced metric at $\psi=100$ over real values of $\psi$. The error band in each case corresponds to the maximal and minimal value obtained respectively when evaluating the $\sigma$ accuracy at different angles. }
  \label{fig:hpsi-don-eta}
\end{figure}

For the main experiment, we have chosen to use uniformly distributed values in the range $0<|\psi|<100$. \figref{fig:hpsi-don-eta} shows the $\sigma$-accuracy achieved by a network with one and two dense hidden layers, respectively. These architectures were chosen as the best-performing ones from a search over several architectures. 
Besides the performance on the training range, the figure shows how well the network extrapolates beyond the training set for larger values up to $|\psi|=1000$.
One can see an improvement in the $\sigma$ accuracies compared to Donaldson's algorithm at the same degree $k$. This improvement is not only present over the range our algorithm was trained on, but extends up to $|\psi|\approx 175$, a factor of 2 beyond the regime used during training.
While the accuracy of the network extrapolation no longer outperforms Donaldson's algorithm beyond this point, it is still a better approximation than an extrapolation of Donaldson's metric at $\psi=100$ over the entire shown range. The latter is shown in the figure as a dashed line.
In summary, the result shows that the machine learning approach can predict very good approximations over a range of values in moduli space. Moreover, it can also extrapolate (within reason) outside the training region.

Our experiments show that this ML approach can outperform Donaldson's algorithm in efficiency (i.e.~the accuracy which can be achieved in a given computing time). The accuracies achieved over the desired range of $\psi$ is strongly dependent on the chosen network architecture, as is the extrapolation behavior. Our primary focus here is showing the feasibility of this ML approach, and we leave further optimization of the architectures to future research.

\subsection{Learning the metric directly}
\label{sec:learningthemetric}
Instead of using a NN to learn the complex structure dependence of the matrix $H$, we can also train a NN to directly learn a functional expression for the CY metric. The value of the metric will depend on the position in the CY manifold as well as on the complex structure. In contrast to the methods presented to learn the K\"ahler potential, we now aim to learn the components of the metric $g$ directly. This has several potential advantages:
\begin{itemize}
\item Instead of the need of predicting $N_k^2$ functions for learning the K\"ahler potential, the NN always only needs to predict the independent components of the metric, i.e.\ $d^2$ real parameters for a complex CY $d$-fold. 
\item In comparison to approaches which use a general ansatz for the K\"ahler potential, learning the metric directly saves two derivatives when evaluating the Monge-Amp\`ere loss.
\end{itemize}
To the best of our knowledge, our experiments are the first to test whether these heuristic differences can be numerically advantageous.

However, there is also a disadvantage as compared to the method discussed in Section~\ref{sec:kaehler}. The metric $g$ is not automatically K\"ahler, nor does it automatically glue nicely across patches of $\bP^{d+1}$. So, in addition to finding a Ricci-flat metric that solves the Monge-Amp\`ere equation~\eqref{eq:MAEquation}, we will need to impose that the K\"ahler and gluing conditions are satisfied. As mentioned previously, the fact that the K\"ahler property is not ensured by construction also allows us to apply this approach to more general (non-K\"ahler) $SU(3)$-structure metrics.

Finding a NN that computes the CY metric for a given point and complex structure then comes down to optimizing the parameters of the NN subject to these three loss components:
\begin{equation}
 \mathcal{L}=\lambda_1 \mathcal{L}_{\rm MA}+\lambda_2 \mathcal{L}_{\rm dJ}+\lambda_3 \mathcal{L}_{\rm overlap}~,
 \label{eq:totalLoss}
\end{equation}
where the optimal weighting $\lambda_i$ of these losses have to be chosen experimentally in a hyperparameter search.

Experimentally, we have found that it is beneficial to start near a solution which satisfies the overlap conditions approximately. While we do not know the CY metric, we know several metrics that are K\"ahler and glue nicely on the CY manifold: the Fubini-Study (FS) metric on the ambient $\bP^{d+1}$ pulled back to the CY space. Since this provides a promising starting point in the sense that out of the three consistency conditions only the Ricci-flatness criterion needs to be optimized, we will start our network as a small perturbation around the Fubini-Study metric utilizing one of the two ans\"atze:
\begin{equation}
 g^{(1)}=g_{FS}+g_{NN}~,~g^{(2)}=g_{FS}~(1+g_{NN})~.
 \label{eq:nnmetricansatz}
\end{equation}
A priori, it is unclear which type of network works better, and we experimentally determined which approach leads to the best result.

When outputting the information about the metric components we have explored two directions, either to output the real and imaginary parts of the metric directly or the components in the LDL decomposition of the metric
\begin{equation}
 g= L D L^\dagger~.
 \label{eq:ldldecomposition}
\end{equation}
Here, $L$ is a complex lower triangular matrix with $1$'s along its diagonal, while $D$ is a real diagonal matrix. When learning the metric directly, we output the independent components of the metric
\begin{equation}
 N(\vec{z},\psi)=\left(g_{11},g_{22},g_{33},{\rm Re}(g_{21}),{\rm Re}(g_{31}),{\rm Re}(g_{32}),{\rm Im}(g_{21}),{\rm Im}(g_{31}),{\rm Im}(g_{32}) \right).
\end{equation}
In the LDL parametrization, we output the non-determined components of this LDL decomposition:
\begin{equation}
 N(\vec{z},\psi)=\left(D_{11},D_{22},D_{33},{\rm Re}(L_{21}),{\rm Re}(L_{31}),{\rm Re}(L_{32}),{\rm Im}(L_{21}),{\rm Im}(L_{31}),{\rm Im}(L_{32}) \right).
\end{equation}
In the latter case, the determinant is computed as ${\rm det}(g)=\prod_i D_{ii}$, but reconstructing the actual metric requires two matrix multiplications.

\subsubsection{Experiments}
\label{sec:resultsmetric}

Below, we present two experiments to demonstrate that this method of learning the metric directly works and produces results that clearly improve throughout training from our starting point.

In the first experiment, we aim to learn the metric using a single neural network for all patches. This network takes as an input the real and imaginary parts of a point on the CY manifold in homogeneous ambient space coordinates, together with the real and imaginary parts of  $\psi$ and the ambient space coordinates in which the pulled-back metric is expressed. It should be noted that the information about the ambient space coordinates is available implicitly to the NN, since we go to the patch where the largest absolute value of the homogeneous ambient space coordinates has been scaled to one and where we have solved for the coordinate with the largest $|\partial p_\psi/\partial z_i|$. Adding information about which ambient space coordinates have not been scaled to one or solved for only resolves a theoretical ambiguity at a measure zero set of points on the CY manifold where two or more ambient space coordinates have the same (largest) absolute value. We checked that omitting this information actually does not impact training or final accuracies. For concreteness, we only display results for $\psi=10$ here.

In the second experiment, we train the network on points in the interval $0<|\psi|<10$, and we use a separate neural network for each patch. Each network takes as input the real and imaginary parts of the points in affine coordinates for the respective patch, together with the complex structure parameter $\psi$.

For both types of networks, we have performed hyperparameter tuning, as discussed in more detail in Appendix~\ref{sec:hyperparametermetric}. The results shown below are achieved with standard feedforward neural networks that have three hidden dense layers and a dense output layer with $9$ output dimensions. In both cases, we have found that the multiplicative ansatz $g^{(2)}$ from Equation~\eqref{eq:nnmetricansatz} outperforms the additive ansatz.

\begin{figure}[t]
\centering
\includegraphics[height=0.25\textwidth]{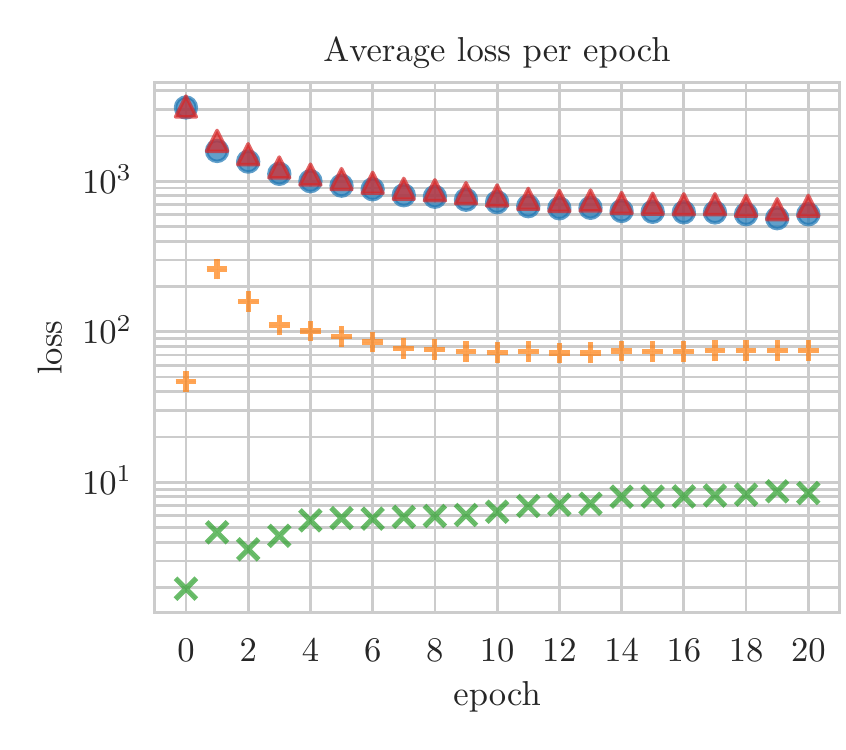}~
\includegraphics[height=0.25\textwidth]{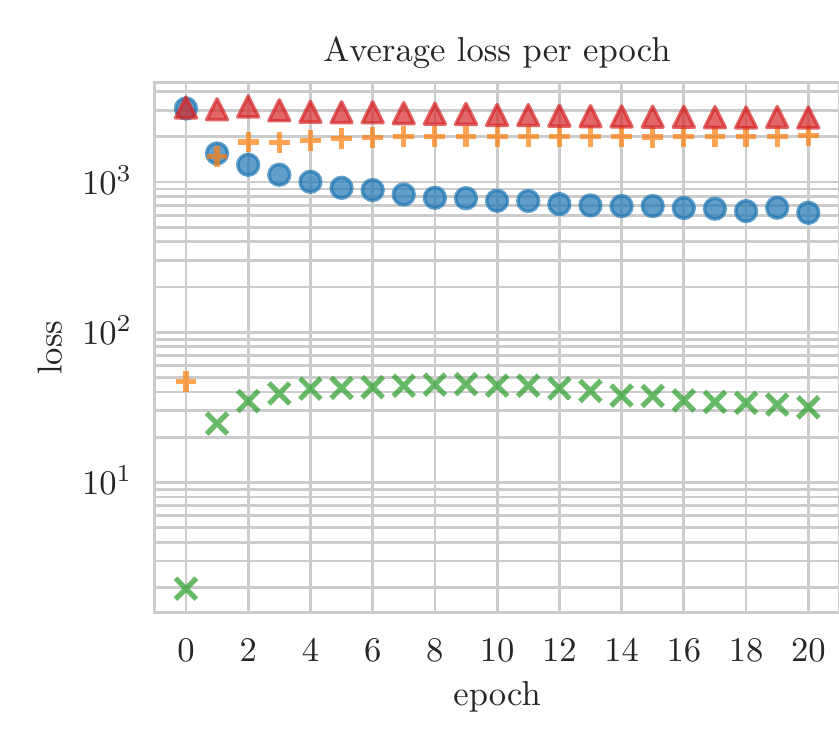}~
\includegraphics[height=0.25\textwidth]{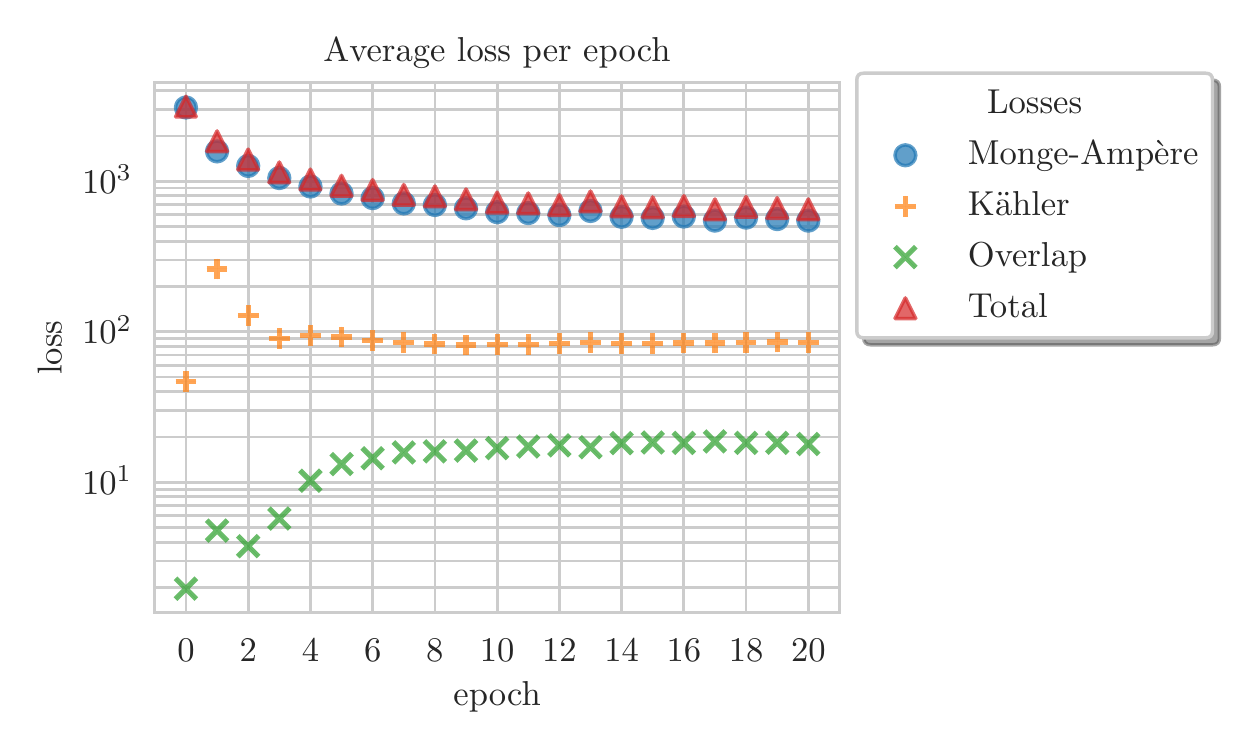}
\caption{Evolution of the training loss during training. \textit{Left:} Optimizing the NN with all three losses. \textit{Middle:} Optimizing the NN without K\"ahler loss (i.e.\ $\lambda_2=0$).  \textit{Right:} Optimizing the NN without overlap loss (i.e.\ $\lambda_3=0$).}
\label{fig:mult-boost-results}
\end{figure}

The results for $\psi=10$ are shown in Figure~\ref{fig:mult-boost-results}. The evolution of the three components that make up the total loss function are plotted on the left. After 20 epochs, the $\sigma$ error measure has gone down from $0.2$ to $0.06$. Note that the $\sigma$ loss can be read off from the Monge-Amp\`ere loss, since they are just proportional with proportionality constant batch\_size$\times\lambda_1=9000$. This flatness accuracy is the same level that is reached with Donaldson's algorithm for $k=6$. We have observed that including more training points (which are easily obtainable) improves the accuracies, and we expect this trend to continue. Note that the initial points at zero epochs provide an approximate comparison to the performance of the Fubini-Study metric, perturbed by a (small) random permutation of the initialized but untrained NN. We also trained the NN with setting $\lambda_2$ (middle) and $\lambda_3$ (right) to zero; in other words, we do not optimize the NN to solve the K\"ahler condition and the overlap condition, respectively. Interestingly, we find that nevertheless, these losses, even if they were not being optimized for in the multiplicative ansatz, do not blow-up. They increase by a factor of 15 and 2.5 respectively when compared with the loss at the end of training where they are included in the optimizer.

\begin{figure}[t]
\centering
  \includegraphics[width=0.32\textwidth]{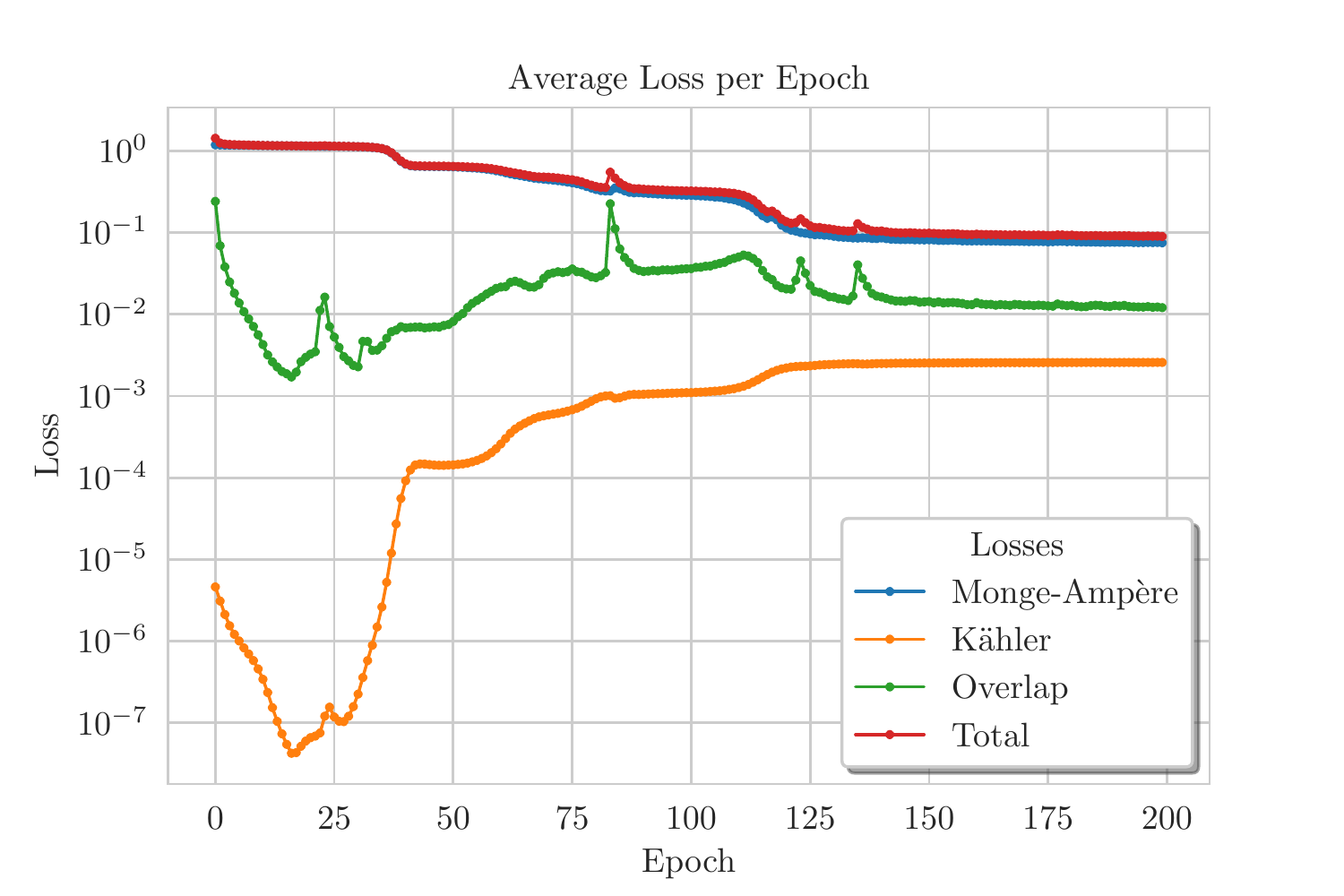}
   \includegraphics[width=0.32\textwidth]{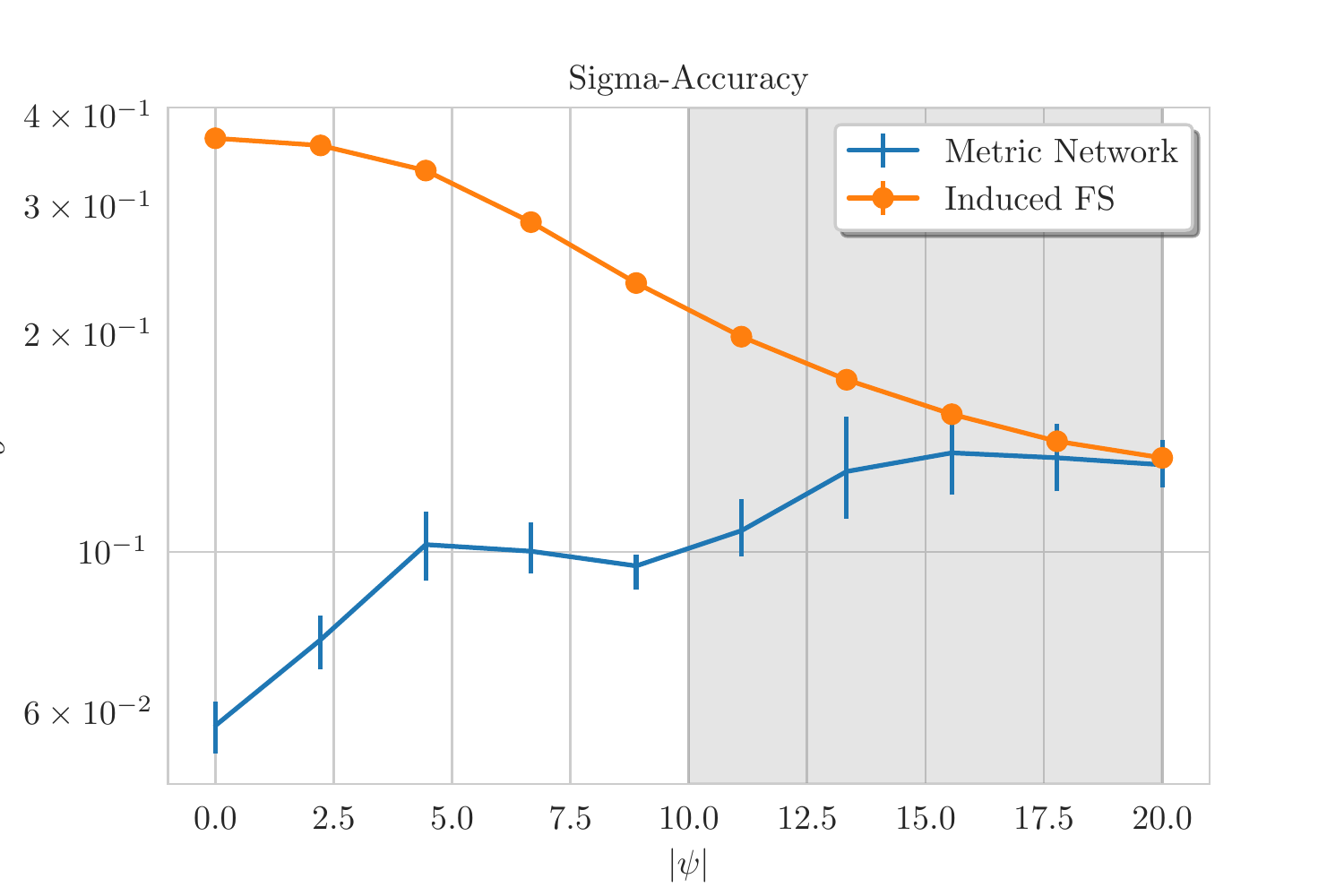}
   \includegraphics[width=0.32\textwidth]{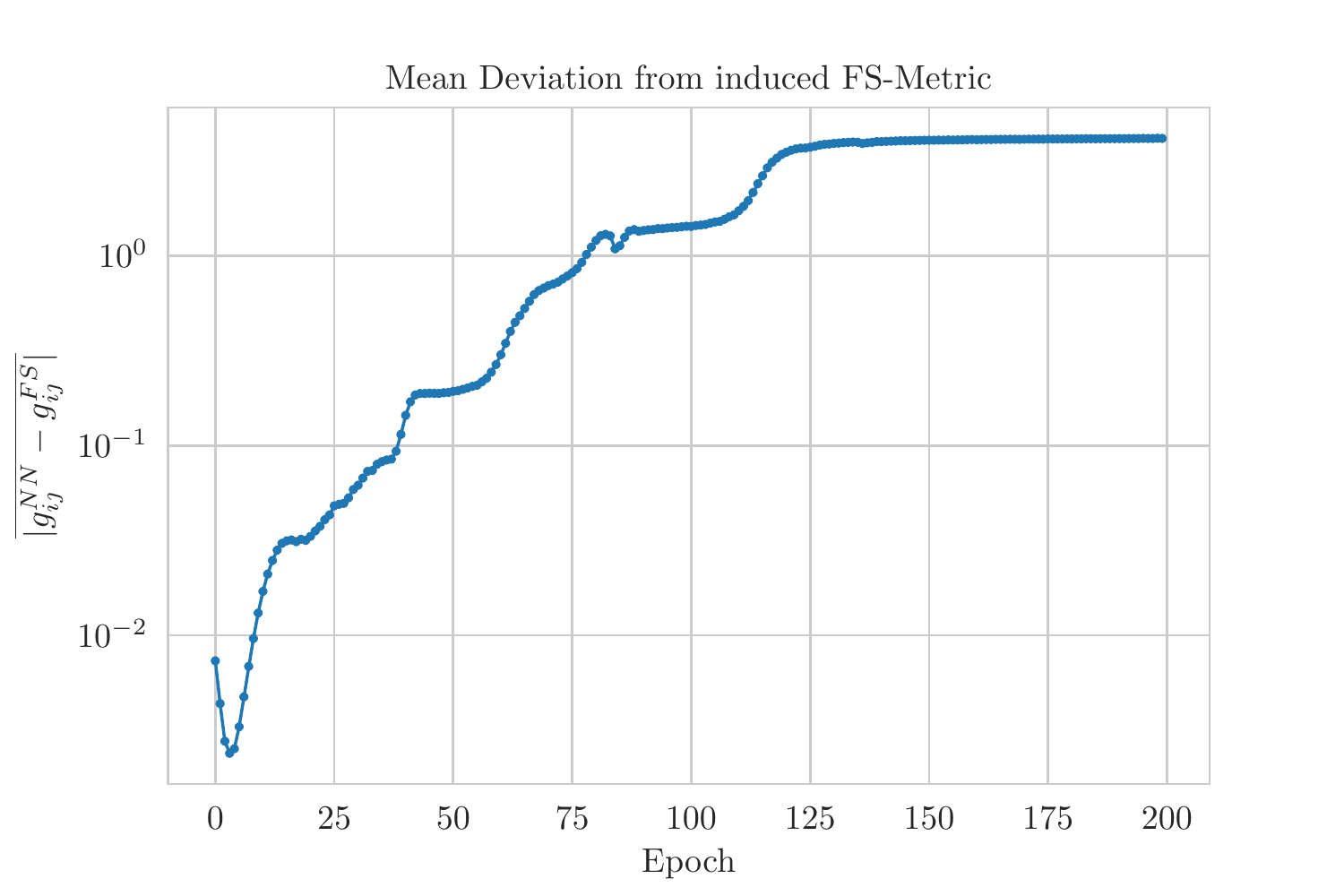}
  \caption{{\bf Left:} Evolution of the validation loss components for our network trained on the range $0<|\psi|<10$. The loss components are averaged over the respective patch-networks. {\bf Middle:} $\sigma$ accuracy of our network in comparison with the induced Fubini-Study accuracy. The data points indicate the mean value of the performance, and the error bars show the minimal and maximal accuracy on different patches. (We have multiple networks which are trained here.) This average is also over four different angles $\theta$ for $\psi=re^{i\theta}$, namely $0,\pi/2,\pi$, and $3\pi/2$. The networks were trained on values of $|\psi|<10$, and the performance in the range $10<|\psi|<20$ is obtained from extrapolation beyond the training set. {\bf Right:} Deviation from the induced Fubini study metric during training, averaged over each patch-network and across the $\psi$ values used for training.}
\label{fig:training}
\end{figure}

In the second example, we learn the LDL-components of the metric, and we have a separate network for each of the five coordinate patches. The training evolution is shown in Figure~\ref{fig:training}, where we see that the $\sigma$ accuracy is improving during training. We observe that the individual networks exhibit jumps in the Monge-Amp\`ere loss at different times, which occur in close vicinity to increases in the overlap loss. In contrast to the previous implementation, we observe that when we do not include the overlap loss condition the overlap conditions are severely violated. Since the first approach has a single neural network for all patches, the difference between the two experiments comes down to the fact that the network in the first experiment has shared weights between the patches, while the NNs in the second experiment have separate weights that are, however, simultaneously optimized.

We want to point out that the magnitudes of the losses in Figures~\ref{fig:mult-boost-results} and~\ref{fig:training} should not be compared directly, as the architectures, the normalization, and the points where the networks were evaluated are different for both experiments. For a rough estimate of the relative scaling of the losses, one can look at the beginning of the training, since both architectures start close to the induced Fubini-Study metric. Alternatively, one can convert the Monge-Amp\`ere losses for the first experiment to $\sigma$ accuracies (as explained above) and then compare the result to the $\sigma$ accuracies for the second experiment, which are plotted in the middle of Figure~\ref{fig:training}.

As a final comment, it should be noted that for the CYs considered in this paper, the K\"ahler class is fixed by the volume as measured by $\int_X J^3$. This can be scaled out and absorbed in $\int_X\Omega\wedge\bar\Omega$. By fixing $\Omega$ from our construction~\eqref{eq:Omega} and by setting $\kappa=1$ in the loss~\eqref{eq:sigma}, we fix the K\"ahler class. In more general setups with $h^{11}>1$, this is not the case. Then, one can simply integrate the pull-back of the numerical K\"ahler form over an appropriate basis of intersections of pairs of divisors. With knowledge of the intersection numbers, one can then pick out the coefficients of the duals of divisor classes in $J$. If one wants to fix the K\"ahler class, one can easily add the K\"ahler class as input to the NNs and enforce it by computing these integrals and adding them to the loss. Alternatively, one can leave the K\"ahler class unspecified, but then one needs to be careful with the Monte Carlo integration: since the weights depend on the K\"ahler class, they need to be re-evaluated with respect to the K\"ahler class of the current, approximate metric after each update.

\section{CY metrics with \textit{SU(3)} structure}
\label{sec:NN-torsion}

In this section, we will turn our attention to metrics that are associated to more general $SU(3)$ structures than the CY metrics with $SU(3)$ holonomy. We will need a small amount of background on $SU(3)$ structures in general and how they appear in string theory compactifications in particular. Here, we will quickly summarize the needed material following the notation and conventions of~\cite{Larfors:2018nce}, before proceeding to discuss how one can set up numerical approaches to finding the associated metrics.

An $SU(3)$ structure on a real six-dimensional manifold can be specified by two nowhere vanishing forms. These are a real two-form $J$ and a complex three-form $\Omega$ satisfying the following algebraic relations.
\begin{eqnarray} \label{algSU3}
J \wedge J \wedge J = \frac{3}{4}i \Omega \wedge \overline{\Omega} \;\;,\;\; J \wedge \Omega=0
\end{eqnarray} 
Given the pair of forms $(J,\Omega)$ obeying the algebraic conditions above, the nature of the resulting $SU(3)$ structure is encoded in five torsion classes. These are determined in terms of the the exterior derivatives of $J$ and $\Omega$,
\begin{eqnarray} \label{diffSU3}
dJ &=& -\frac{3}{2}\textnormal{Im}(W_1 \overline{\Omega}) + W_4 \wedge J + W_3 \\ \nonumber
d\Omega &=& W_1 J \wedge J + W_2 \wedge J + W_5 \wedge \Omega\;,
\end{eqnarray}
together with the conditions $W_3 \wedge J = W_3 \wedge \Omega = W_2 \wedge J \wedge J =0$, which make the above decomposition of $dJ$ and $d\Omega$ unique.
Frequently in what follows it will be useful to have straight forward formulae for extracting the torsion classes given $J$ and $\Omega$. They read:
\begin{eqnarray} \label{extract}
W_1 = -\frac{1}{6} i \Omega \lrcorner dJ = \frac{1}{12} J^2 \lrcorner d\Omega \;\;,\;\; W_4=\frac{1}{2}J \lrcorner dJ \;\;,\;\; W_5 =-\frac{1}{2} \Omega_+\lrcorner d\Omega_+~.
\end{eqnarray}
Here, we use subscripts of $\pm$ to indicate real and imaginary parts, and the symbol $\lrcorner$ denotes contraction with indices being raised with the metric. Given the expression for three of the torsion classes in~\eqref{extract}, the other two classes $W_2$ and $W_3$ can be trivially obtained from~\eqref{diffSU3}. And given the data $(J,\Omega)$ of an $SU(3)$ structure, one can easily reconstruct the associated metric as $g_{mn}={\cal J}_m^{\;l}J_{ln}$ where ${\cal J}_m^{\;l}$ is the almost complex structure determined by the three-form $\Omega$.

\vspace{0.2cm}

As mentioned in the introduction, a wide variety of $SU(3)$ structures appear in the subject of string compactifications. By far the most widely studied case is that were all of the torsion classes vanish: this reduces to the Ricci-flat CY manifolds that have been the focus of previous section. However, this special case is frequently studied simply for computational ease, and many other possible torsion classes are of interest. In this paper, we will present just one illustrative example: the general constraints on the torsion classes of the Strominger-Hull system that are required for an ${\cal N}=1$ four-dimensional Minkowski vacuum in heterotic string theory~\cite{Strominger:1986uh,Hull:1986kz,LopesCardoso:2002vpf}. In that case, the requirement for a good supersymmetric vacuum can succinctly be stated as follows:
\begin{eqnarray} \label{elstrom}
W_1=W_2=0\;\;, \;\;  W_4 = \frac{1}{2} W_5= d\phi \;\;,\;\; W_3\;\; \textnormal{arbitrary}~.
\end{eqnarray}
Here $\phi$ is the heterotic dilaton. In this section, as an example of how machine learning techniques can be used to numerically find metrics associated to non-Ricci-flat $SU(3)$ structures, we will generate metrics associated to structures of this form.

\subsection{Learning an ansatz}

One of the more difficult issues in numerically searching for an $SU(3)$ structure is to ensure that the forms $J$ and $\Omega$ are globally well-defined and nowhere vanishing. One approach to addressing this issue is to impose an ansatz which enforces such behavior from the outset. As an example for this, we will consider a generalization of the ansatz that was considered in~\cite{Larfors:2018nce}. This approach to machine learning $SU(3)$ structure metrics is somewhat similar in spirit to Section~\ref{sec:predictingHdirectly}. It is important to note that in most of the discussion that follows, we will consider the case where the moduli have been fixed to a specific value. 

The ansatz we will consider will provide (torsional) $SU(3)$ structures on CY three-folds described as a complete intersection in products of projective spaces (CICYs). We can describe such a manifold in terms of a configuration matrix.
\begin{eqnarray} \label{configmat}
\left[\begin{array}{c|ccc} \bP^{n_1} &q^1_1& \ldots& q^1_K \\ \vdots& \vdots&\vdots&\vdots \\  \bP^{n_m}& q^m_1 &\ldots& q^m_K \end{array} \right]~.
\end{eqnarray}
Such a manifold is the common solution set of $K$ homogeneous equations in an ambient space $\bP^{n_1} \times \ldots \times \bP^{n_m}$. Each column of $q$'s in (\ref{configmat}) denotes the homogeneous multi-degree of one of the $K$ defining equations in the coordinates of the ambient space factors. Clearly, the complex dimension of such a manifold is $\sum_{i=1}^m n_i -K$. The condition that the first Chern class vanishes can be satisfied by insisting that $\sum_r q^i_r =n_i+1$ for all $i$.

On such a manifold, we make the following ansatz
\begin{eqnarray} \label{ans2}
J=\sum_i^m a_i J_i \;\;,\;\; \Omega=A_1 \Omega_0 + A_2 \overline{\Omega}_0~.
\end{eqnarray}

Here, $J_i$ is the restriction to the CY manifold of an algebraic K\"ahler form for the $i^{\rm th}$ ambient projective space factor (this can be derived from a K\"ahler potential described by~\eqref{eq:generalisedFS}). Meanwhile, $\Omega_0$ is the usual expression for the closed holomorphic three-form associated with the Ricci-flat structure on the CICY~\cite{AtiyahBottGarding,Candelas:1987kf}.
The $a_i$ are $m$ real functions, while $A_1$ and $A_2$ are complex functions. This ansatz becomes the same as that which was used in the analytic work of~\cite{Larfors:2018nce} if we set $A_2=0$ and replace the $J_i$ with the restriction of Fubini-Study K\"ahler forms. We note that including the form $\overline{\Omega}_0$ in the ansatz for $\Omega$ can be important in that it allows us to divorce the almost complex structure of the $SU(3)$ structure being considered from the integral complex structure inherited from the ambient space. We will see this in more detail shortly. 

The benefit of an ansatz such as (\ref{ans2}) is that it automatically ensures that $J$ and $\Omega$ are nowhere vanishing and globally well-defined if the $a_i$ are taken to be everywhere positive and if $A_1$ and $A_2$ are chosen to be nowhere vanishing. In addition, this ansatz automatically defines an $SU(3)$ structure for such choices of the undetermined functions, subject to one further condition. While the second condition in (\ref{algSU3}) is automatic, the first is only satisfied if the following relationship between the functions holds:
\begin{eqnarray} \label{cons2}
|A_1|^2+|A_2|^2=\sum_{i,j,k=1}^m \Lambda_{ijk}a_ia_j a_k
\end{eqnarray}
In this expression, $\Lambda$ is defined via the following equation.
\begin{eqnarray}
J_i \wedge J_j \wedge J_k= \frac{3}{4}i \Lambda_{ijk}\Omega_0 \wedge \overline{\Omega}_0
\end{eqnarray}
Thus the ansatz (\ref{ans2}), subject to the constraint (\ref{cons2}) gives rise to a $SU(3)$ structure, for any appropriate choice of the functions that appear.

Given such an $SU(3)$ structure, we can compute its torsion classes using (\ref{diffSU3}) and (\ref{extract}). We find that 
\begin{eqnarray} \label{tors1}
W_1&=&0 \\\nonumber
W_2 &=& -i \overline{\partial} A_1 \lrcorner \Omega_0 + i \partial A_2 \lrcorner \overline{\Omega}_0 +i\frac{\overline{\partial} (A_1 + \overline{A}_2)}{A_1+ \overline{A}_2} \lrcorner A_1\Omega_0-i\frac{\partial(\overline{A}_1 + A_2)}{\overline{A}_1+ A_2}\lrcorner A_2 \overline{\Omega}_0\\ \nonumber
W_3 &=& \sum_i(da_i-W_4)\wedge J_i \\\nonumber
W_4 &=& \frac{1}{2} \sum_i J_i \lrcorner(da_i\wedge J_i) \\\nonumber
W_5 &=& \frac{\overline{\partial} (A_1 + \overline{A}_2)}{A_1+ \overline{A}_2}+\frac{\partial(\overline{A}_1 + A_2)}{\overline{A}_1+ A_2}~.
\end{eqnarray}
Note that if we set $A_2=0$ we regain the expressions produced in~\cite{Larfors:2018nce} where $W_2=0$ and the form of $W_5$ was simpler. We see again here that the generalization of the ansatz we are introducing does produce a qualitative difference to that which appeared in~\cite{Larfors:2018nce}, even when replacing the $J_i$ with Fubini-Study K\"ahler forms. The almost complex structure which is associated to the $SU(3)$ structures described by the ansatz is no longer strongly linked to that of the $SU(3)$ holonomy structure. As such, it no longer has to be integrable: we can describe non-integrable almost complex structures on the underlying complex manifold in this manner, leading to the non-vanishing $W_2$ in (\ref{tors1}).

\vspace{0.2cm}

Given the ansatz (\ref{ans2}), our goal is to set up a NN which takes as input the real and imaginary parts of a point on the CICY threefold (in terms of homogeneous ambient space coordinates), together perhaps with the real and imaginary parts of some coefficients in the defining relation if such dependence is desired. As an output, the NN should give the $a_i$, $A_1$, $A_2$ and the $H$ parameters appearing in the $J_i$'s, perhaps with some additional ancillary data as we will describe shortly.

In terms of loss functions, several of the requirements that should be imposed are automatically satisfied by (\ref{ans2}). There is no need to have a contribution to the loss function which aims to enforce global well-definedness and non-vanishing, for example, as we did in Section~\ref{sec:learningthemetric}. The ansatz itself guarantees the former, and encoding the $a_i$ and the real and imaginary parts of $A_1$ and $A_2$ as exponentials of real functions would be sufficient to enforce the latter. The result is also guaranteed to be an $SU(3)$ structure given the above discussion, if (\ref{cons2}) holds. We have two options here. We can solve (\ref{cons2}) explicitly for one of the defining functions of the ansatz in terms of the others. Or we can include a contribution to the loss function of the form,
\begin{eqnarray} \label{SU3loss1}
{\cal L}_{SU(3)} = \left\lVert |A_1|^2+|A_2|^2 - \sum_{i,j,k=1}^m \Lambda_{ijk}a_ia_j a_k \right\rVert_n \;.
\end{eqnarray}

\vspace{0.1cm}

The remaining contributions to the loss function would all be concerned with the torsion classes of the $SU(3)$ structure that we are trying to produce. Instead of imposing a loss function trying to enforce K\"ahlerity as in (\ref{Kloss}), one would ask instead that the $W_i$ take a given desired form. What would be required here would depend upon the physical application, with different string constructions placing different constraints upon the torsion classes. As a concrete example, let us discuss the loss functions that would be used if a solution to the Strominger system (\ref{elstrom}) was desired.

We see from (\ref{tors1}) that $W_1=0$ is automatic given our ansatz, and for the Strominger system $W_3$ is arbitrary so that we do not need to include these quantities in any loss function. This just leaves us with $W_2=0$ and $2 W_4= W_5 = d \phi$ as constraints to consider. Combining (\ref{diffSU3}) and (\ref{extract}), together with the condition $W_1=0$, we obtain
\begin{eqnarray}
W_2 \wedge J= d \Omega+ (\frac{1}{2} \Omega_+ \lrcorner d \Omega_+)\wedge \Omega~.
\end{eqnarray}
The condition $W_2=0$ can therefore be enforced by including the loss function contribution
\begin{eqnarray}  \label{SU3loss2}
{\cal L}_{W_2} = \left\lVert d \Omega+ (\frac{1}{2} \Omega_+ \lrcorner d \Omega_+)\wedge \Omega \right\rVert_n~.
\end{eqnarray}

The final set of conditions $2 W_4= W_5 = d \phi$ is slightly less straightforward given that we currently do not know, in any realistic application, what the profile for the heterotic dilaton $\phi$ would be. This leads us to include $\phi$ as part of the output of the NN: this is an example of the extra ancillary data that can sometimes be required in the output that was mentioned above. Given the expressions for $W_4$ and $W_5$ in (\ref{extract}) we then add the following contributions to the loss function:
\begin{eqnarray} \label{SU3loss3}
{\cal L}_{W_4} &=&  \left\lVert J \lrcorner dJ - d \phi \right\rVert_n \\ \nonumber
{\cal L}_{W_5} &=& \left\lVert -\frac{1}{2} \Omega_+ \lrcorner d \Omega_+ - d\phi \right\rVert_n
\end{eqnarray}

Combining the contributions in (\ref{SU3loss1}), (\ref{SU3loss2}) and (\ref{SU3loss3}), we then arrive at the following total loss function, in the case where we are interested in $SU(3)$-structure solutions to the Strominger system:
\begin{eqnarray}
{\cal L}_{\textnormal{Strominger}} = \gamma_1 {\cal L}_{SU(3)} + \gamma_2 {\cal L}_{W_2} + \gamma_3 {\cal L}_{W_4} + \gamma_4{\cal L}_{W_5}
\end{eqnarray}
Here, the $\gamma_i \in \mathbbm{R}^+$ allow us to weight the various conditions being imposed differently, analogously to the $\lambda$'s in (\ref{eq:totalLoss}). In the case where one solves (\ref{cons2}) analytically, one would of course set $\gamma_1=0$. Clearly, analogous loss functions could be set up for the constraints placed upon torsion classes by other string compactifications.

\subsubsection{An example} \label{aeg}

Running a full analysis of an ansatz of the type described above is too complex for a first attempt at using machine learning techniques to learn $SU(3)$-structure metrics. (Indeed, we believe this is the first work on numerical $SU(3)$-structure metrics of any kind in the physics literature). As such, instead of providing an explicit example in this sub-section, we will defer providing sample computational results until the next. However, there is one last issue that we should address before moving on to the subject of directly learning the $SU(3)$-structure metric. In developing NN's to describe $SU(3)$ structures, there is a question as to how to evaluate the trustworthiness of the results. Numerical methods for constructing Ricci-flat metrics on CY manifolds benefit from several notable advantages over those aimed at producing more general structures. One of these is that existence theorems guarantee that a solution to the system exists. This is important as it shows that the numerical approximations that are being obtained are close to full solutions to the system, rather than just being metrics which approximate the desired properties in a system which admits no exact solution. In particular, the method utilizing extremization of an energy functional~\cite{Headrick:2009jz,Cui:2019uhy} can rest on Yau's theorem~\cite{yau}, and Donaldson's approach~\cite{Donaldson:2005mat,Douglas:2006hz,Douglas:2006rr,Braun:2007sn} can use certain results pertaining to balanced metrics and the algebraic ansatz for the K\"ahler potential~\cite{Donaldson:2005mat,Donaldson2}. In the case of more general $SU(3)$ structures, there are, to our knowledge, no such existence theorems available.

To combat this issue, we will show that the numerical results that we will present in the next sub-section approximate an explicitly known $SU(3)$-structure solution with torsion on the quintic CY threefold~\cite{Larfors:2018nce}. For this solution, the authors of~\cite{Larfors:2018nce} give the following expressions for the functions appearing in (\ref{ans2}) where $J$ is the K\"ahler form derived from the Fubini-Study K\"ahler potential:
\begin{eqnarray}
\label{eq:ansatz-example}
&& a_1 = \frac{1}{\pi^3} \frac{|\nabla p|^2}{\sigma^4} \;\;,\;\; A_1= a_1^2 \;\;,\;\; A_2= 0~, \\ \nonumber
&&\textnormal{where} \;\;\;  \sigma= \sum_{a=0}^4|X_a|^2~.
\end{eqnarray}
In this expression, $p$ is the defining relation of the quintic hypersurface and the $X_a$ are the homogeneous coordinates on $\bP^4$. In addition, the authors take $J_1$ to be the Fubini-Study K\"ahler form of $\bP^4$ restricted to the quintic CY threefold, rather than the more general algebraic K\"ahler potentials considered in (\ref{ans2}). These choices lead, from (\ref{tors1}), to torsion classes
\begin{eqnarray}
\label{eq:torsion-example}
W_1=W_2=W_3=0 \;\;,\;\; W_5 = 2 W_4 = 2 d (\ln a_1 )~.
\end{eqnarray}
Comparison with~\eqref{elstrom} shows that these choices indeed lead to a solution to the torsion class constraints that arise from considering the Strominger system of heterotic string theory. To show that the methodology being proposed in this section for machine learning $SU(3)$-structure metrics is viable, we will, in the next sub-section, show that the techniques being implemented can correctly reproduce this known solution.

\vspace{0.2cm}

\subsection{Learning the \textit{SU(3)} structure directly}

Starting from an ansatz such as (\ref{ans2}) for an $SU(3)$ structure, as we did in the last section, has many advantages. The resulting structure is automatically globally well defined. It is also automatically an $SU(3)$ structure if we choose to solve (\ref{cons2}) analytically. Nevertheless, just as considered in Section~\ref{sec:learningthemetric} for the CY case, one could try and learn the K\"ahler form (and threeform) of an $SU(3)$ structure directly rather than leaning on an ansatz. Such an approach, while much more ambitious, clearly has potential advantages. For example, the ansatz (\ref{ans2}) we have provided relies on the existence of known nowhere vanishing forms on the space on which it is defined -- restricting the possible manifolds to which analogous techniques can be applied. In addition, we can see from equation (\ref{tors1}) that the ansatz (\ref{ans2}) is constrained in the forms of torsion classes it can give rise to.

One possible strategy would be to take as inputs to a NN the real and imaginary parts of points on some algebraic variety, and as outputs the components of the real two-form $J$ and the real and imaginary parts of the components of the complex three-form $\Omega$ at those points. The global well definedness of the forms could be imposed using loss functions contributions similar to (\ref{transloss}) and simple contributions imposing the algebraic conditions (\ref{algSU3}) can be implemented in a trivial manner. Contributions to the loss function guaranteeing that $J$ and $\Omega$ were nowhere vanishing, perhaps by constraining the eigenvalues of $J$ at each point and the contraction of $\Omega$ with its complex conjugate, would also have to be included.

An important advantage to such an approach to numerically determining $SU(3)$ structures for string compacitifications would be the ability to `choose' any set of torsion class constraints by appropriate choices of loss functions. In the analytic approaches to $SU(3)$ structures that have been applied to date, one first makes an ansatz for the geometry involved and then computes the torsion classes that can be achieved. This is a shooting problem in that there is no guarantee that a given choice of ansatz may be capable of reproducing the torsion classes necessary for a given type of string compactification. In an approach such as that being discussed here, analogues of (\ref{SU3loss2}) and (\ref{SU3loss3}) could be used to obtain any pattern of torsion classes desired, assuming that such a pattern is possible on the manifold under consideration.

To illustrate this approach to numerically determining $SU(3)$-structure metrics, we will provide an explicit example rather than outlining general formalism. The form a general approach would take is rather clear given the above discussion, and it is perhaps useful at this stage to present a concrete result. Rather than attempting to learn both forms of the $SU(3)$ structure in what follows, we will specify $\Omega$ and attempt to learn $J$. Such a simplification has two benefits. First, it makes this initial foray into such work simpler. Second, in doing so we will show that we can guide the system to learn the known example of an $SU(3)$ structure obtained in~\cite{Larfors:2018nce} and repeated in Section~\ref{aeg}. This is important, since reproducing a known solution gives us more confidence in the methods being espoused, given the lack of existence theorems in this setting.

In more detail, we fix a three-form for the $SU(3)$ structure $\Omega$ by using (\ref{ans2}) and~\eqref{eq:ansatz-example}. In that case, the torsion class $W_5=2 d(\ln a_1)$ is fixed by~\eqref{extract} and we have $W_1=W_2=0$. If we wish to look for torsion classes compatible with the Strominger-Hull system, then this also fixes $W_4=\frac{1}{2}W_5$ via (\ref{elstrom}). In order to try and reproduce the known solution of Section~\ref{aeg}, we will look for a case where $W_3=0$. In general, we are not guaranteed to find a solution with $W_3=0$ and indeed we could leave this torsion class as an output of the NN rather than specifying its value. However in the case at hand, requiring its vanishing will allow us to verify the validity of our techniques by recovering the known solution of Section~\ref{aeg}. We then have a complete specification of the torsion classes desired and can attempt to learn the two form of the $SU(3)$ structure $J$.

We will need several contributions to the loss function. First we implement a loss of the form
\begin{eqnarray} \label{su3loss}
{\cal L}_{SU(3)} = \left\lVert 1 +i \frac{4}{3} \frac{J \wedge J\wedge J}{\Omega \wedge \overline{\Omega}} \right\rVert_n
\end{eqnarray}
in order to impose the first of the algebraic conditions defining the $SU(3)$ structure appearing in (\ref{algSU3}). Note that given the index structure we will impose on $J$ and the form of $\Omega$ being taken, the second of the constraints in that equation are automatic. This is the same loss as appeared in (\ref{eq:algconstloss}), given a different name as we are no longer searching for a Ricci-flat metric. It is a useful fact that this loss function also enforces the nowhere vanishing condition on $J$, given the nowhere vanishing nature of the expression being used for $\Omega$. We also need to impose the transition loss~\eqref{transloss} which also remains unchanged from the CY case.  In order to impose the torsion class constraints discussed above, we can use equation~\eqref{diffSU3}. For our torsion classes, the condition simply becomes
\begin{align}
dJ = W_4\wedge J\,.
\end{align}
Hence, we will use the loss
\begin{align}
\mathcal{L}_{W_4}'=|| dJ-d\ln a_1\wedge J ||_n\,,
\end{align}
which closely resembles the K\"ahler loss~\eqref{Kloss}.

\begin{figure}[t]
\centering 
\includegraphics[width=.7\textwidth]{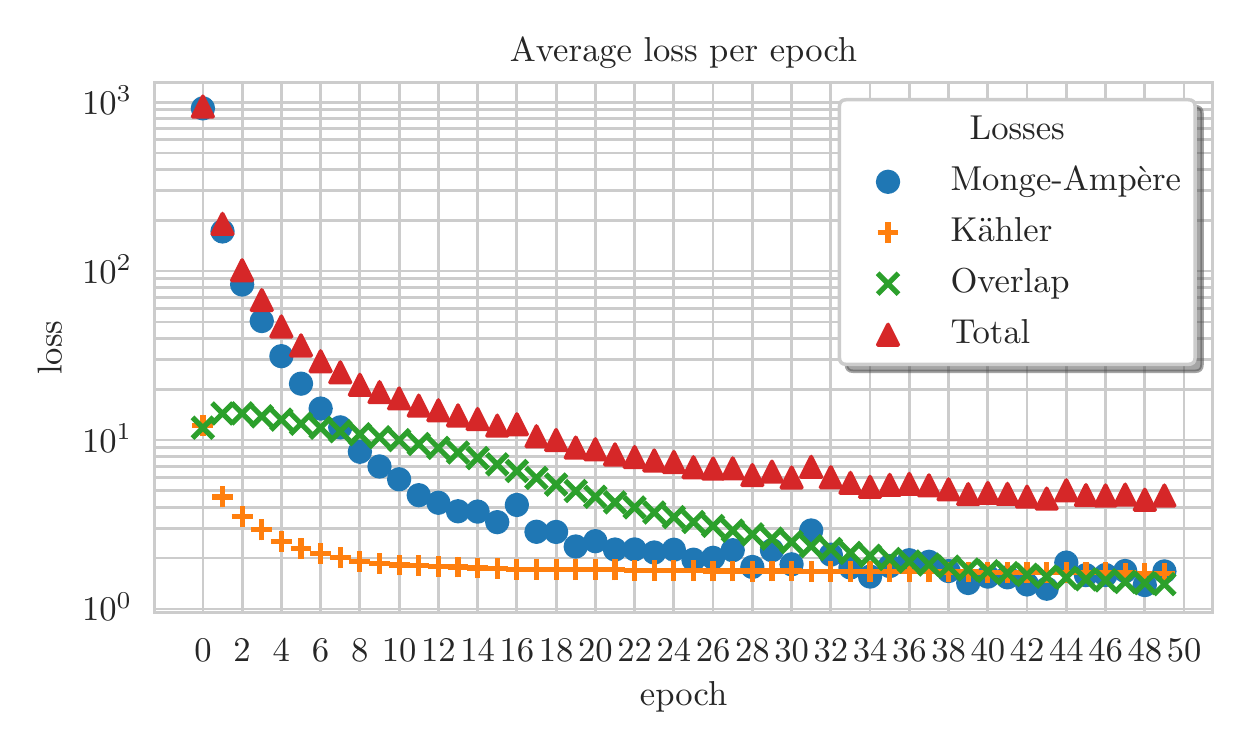}
\caption{\label{fig:su3loss} Change in loss during training for the $SU(3)$-structure example.}
\end{figure}

We will use the same example as for the CY metrics in earlier sections, i.e.\ the quintic with one parameter $\psi=10$. We also leave all other hyperparameters unchanged; in particular, we choose the weight factor $\gamma_1$ of the contribution to the $SU(3)$ loss function to be 10, all other $\gamma_i$ to be one, and set $n=1$ (so that we are using the L$_1$ norm for the losses and not weighting outliers disproportionately strongly). We use multiplicative boosting from the Fubini-Study metric. Figure~\ref{fig:su3loss}  shows how the losses change over the course of training.
As a measure for how much the metric improves during training as compared to the Fubini-Study metric, we compute the equivalent of the $\eta$ error measure, i.e.\ the departure from the Monge-Ampere equation averaged over all points on the manifold in the test set:
\begin{align} \label{fromtheloss}
\langle\eta_{SU(3)}\rangle=\frac{1}{N_\text{pts}}\sum_{i=1}^{N_\text{pts}} \left|\left. 1-\frac{3i}{4}\frac{\Omega\wedge\bar\Omega}{J^3}\right|_{p_i}\right|\,.
\end{align}
We find that if we set the NN to zero, i.e.\ use the FS metric as the lowest order approximation to the $SU(3)$-structure metric, we get $\langle\eta_{SU(3)}(g_{FS})\rangle\approx 40'000$. In contrast, the metric after training gives $\langle\eta_{SU(3)}(g_{NN})\rangle\approx 1.2$, i.e.\ an improvement of 5 orders of magnitude. 

\vspace{0.1cm}

The error measure (\ref{fromtheloss}) is closely related to the loss function (\ref{su3loss}). In addition, this quantity is only a measure of how close we are to {\it some} $SU(3)$ structure. It does not demonstrate that we are correctly approximating the analytic example described in Section~\ref{aeg}. In order show that our numerics are approaching this known solution we wish to consider an error measure of the following form.
\begin{eqnarray} \label{errorthingy}
{\cal E}_{\textnormal{known}} = || g_{\textnormal{numeric}} - g_{\textnormal{known}}||_n
\end{eqnarray}
In this expression $g_{\textnormal{numeric}}$ is the output of our trained NN and $g_{\textnormal{known}}$ is the known solution computed from the quantities given in Section~\ref{aeg}. In fact, some caution is required here as even if the numerical results were approaching the analytic expression, the two could be related by a non-trivial coordinate transformation. If such a coordinate transformation is to preserve the form of the quintic polynomial (\ref{eq:CYHypersurfaces}), then it must be linear. Additional constraints are placed upon this transformation by the requirement that it preserve $\Omega$, which is the same for the numerical and exact solutions. Imposing these two constraints on the set of possible coordinate transformation provides us with a small list of possibilities that must be considered, and in considering (\ref{errorthingy}) we choose the transformation that minimizes its value.

Proceeding in this manner, we obtain a measure of how accurately our NN is reproducing the analytic solution of Section~\ref{aeg}. If we choose $n=1$, we find that ${\cal E}_{\textnormal{known}}$ goes from $0.511$ for the Fubini-Study metric to $0.025$ for the output of the NN. Moreover, choosing higher values of $n$ makes the improvement even more notable - showing that the numerical result has fewer outlying regions that are far from the desired solution. For example, if we choose $n=2$ then we obtain values of $0.59$ and $0.017$ respectively. Thus we find that the machine learning techniques described in this section are indeed capable of reproducing known results for $SU(3)$ structures on six-manifolds. This gives us confidence that such techniques can be useful in this arena going forward.

\vspace{0.2cm}

One can imagine many long term goals of the approach to obtaining explicit $SU(3)$ structure metrics discussed in this section. For example, one could in principle add contributions to the loss function designed to ensure that there are no small cycles anywhere in the target geometry (an issue common to all known $SU(3)$ structure solutions with non-trivial torsion to date). Even the most basic implementation of this approach is beyond the scope of the current paper, however, and we leave the exploration of such possibilities to future work.

\section{Conclusions and future directions}
\label{sec:Conclusions}

CY geometries play an important role in string compactifications. However, the fact that no explicit, analytic CY metrics are known has formed a substantial barrier to progress in a wide range of physical applications. As a result, the need for numerical approximations has been long-standing. In this work, we have demonstrated that the techniques of ML can serve as an important addition to this literature, producing results on par with or surpassing those obtained from methods such as the Donaldson algorithm and energy minimization, while at the same time naturally including complex structure moduli dependence. In particular, the techniques presented in the previous sections provide both certain quantitative and qualitative improvements on the prior state of the art. The most significant qualitative advances being that machine learning techniques allow us to effectively study moduli dependence of CY metrics (something very difficult to achieve with the Donaldson algorithm for example, which is formulated at a single point in the CY moduli space) and importantly, to move away from the complex/K\"ahler regime entirely, by approximating Ricci-flat but non-K\"ahler metrics for manifolds of special structure.

The key results of this work include the following:
\begin{itemize}
\item We have demonstrated that ML is a viable approach to finding Ricci-flat metrics in the case of $SU(3)$-holonomy and $SU(3)$-structure manifolds. Comparing to existing methods, we find that networks with relatively few dense layers converging to the algebraic metrics outperform Donaldson's algorithm in terms of efficiency (i.e.~with respect to the achieved accuracy given a certain runtime, cf.~Figure~\ref{fig:hpsi-don-eta}). We also find that our metrics generalize well beyond the range they have been trained on. In general the runtime of all networks is very reasonable and our results can be obtained on standard desktop CPU or GPU systems.

\item We have presented the viability of two distinct approaches to approximating a CY metric: 1) learning the K\"ahler potential and 2) directly learning the metric (Figures~\ref{fig:mult-boost-results} and~\ref{fig:training}). This latter approach is a crucial step away from past approaches (which were, by construction, tied to K\"ahler geometry) and the first to be generalizable to metrics for $SU(n)$ structure.

One additional difficulty arises for our directly learned metrics, namely that the loss on the overlap and for the K\"ahler condition is non-vanishing. Pragmatically, we observe that the loss can be kept at a small order compared to what we have started out with, while at the same time the Monge-Amp\`ere-loss is changed by an order of magnitude. We hence consider these solutions as non-trivial approximations for Ricci-flat metrics. This allows us to also search for general solutions with $SU(3)$-structure. We demonstrate for the first time that NNs can find such solutions (Figure~\ref{fig:su3loss}) by reproducing the known, exact results of~\cite{Larfors:2018nce}.

\item We have demonstrated that ML can shine light on previously difficult to determine moduli dependence of CY metrics. In particular, we have applied Donaldson's algorithm to obtain expressions for the CY metric at different points in complex structure moduli space and then trained a NN to learn from that the CS moduli dependence (Figures~\ref{fig:learnHrandom} and~\ref{fig:learnHbox}).

\item Within the context of $SU(3)$-structure solutions, our methods have a potentially important flexibility in that it is possible to approximate a metric \emph{given an explicit choice of torsion classes}. This is in contrast to most other available methods of generating $SU(3)$-structure solutions, which often fix the torsion classes. This flexibility could prove useful in applications within string model building.
\end{itemize}

There are many possible directions in which this work could be extended or applied in the future. Beginning with CY metrics, it is clear that our approach could be readily extended to more general algebraic varieties. For concreteness in the present work, we focused on the quintic one parameter hypersurfaces. However, our architectures can easily accommodate the additional complexity of complete intersection manifolds in more general ambient spaces. Towards this end, we find it encouraging that our algebraic metrics for $k=6$ are optimizing all components of $H$ rather than just non-vanishing components due to symmetry constraints (which have been heavily employed in previous work to ensure that algorithms can actually finish in finite time). In a related spirit, we view the metrics with $SU(3)$ structure studied here as a proof of concept that ML methods are capable of producing non-K\"ahler results. Clearly, it would be of interest to continue such investigations into more general classes of $SU(3)$-structure metrics, or indeed to any special structure manifold. As one particular example, applying ML techniques to metrics for manifolds with $G_2$ holonomy/structure could potentially provide interesting new classes of examples, where existing examples/constructions are scarce. 

It is clear that are a number of natural and very related geometric applications for these tools and the approximate CY metrics we have generated. Many string compactifications involve additional geometric data in the form of slope-stable vector bundles, fluxes, or special sub-cycles (including Special Lagrangian subvarieties of CY 3-folds). The techniques we have developed here could readily be extended to learn these associated structures -- for example the associated Hermitian-Yang-Mills connection on a slope poly-stable vector bundle (something that has already been attempted via the Donaldson Algorithm~\cite{Wang,Douglas:2006hz,Anderson:2009nt,Anderson:2011ed}). Lastly, we could use the approximate metrics generated here to probe theoretically expected structure. This could include decompositions of the metric into fiber/base components in the case of elliptic or K3 fibrations, or in the large complex structure limit one should be able to see that any CY manifold is a $T^3$ fibration according to the SYZ conjecture~\cite{Strominger:1996it}.
 
Finally, our primary goal in beginning this study was the hope that these tools will be of use in applications to string phenomenology and the study of the string swampland. As mentioned previously, canonically normalized kinetic terms are needed to determine particle masses/excitations in string vacua, and for this the explicit metric must be known (see e.g.~\cite{Braun:2008jp}). These masses, together with their moduli dependence, play an important role in the recent discussion of the string swampland, especially in the distance conjecture~\cite{Ooguri:2006in}. Finally, the moduli dependence of the metric will also play a vital role in the quest for moduli stabilization. We hope to turn to some of these open questions in future work.

\section*{Acknowledgments}
We would like to thank Chris Beasley, Michael Douglas, Koji Hashimoto, Andre Lukas and Dieter L\"ust for helpful discussions. We are very grateful to various conferences and workshops in recent years which allowed us to discuss in various combinations about this research and to present preliminary results of this work. In particular this includes: MITP program on {\it String Theory, Geometry and String Model Building}, SCGP programs on {\it Geometry and Physics of Hitchin Systems} and {\it Neural Networks and Data Science Revolution}, String Phenomenology conferences in 2018 and 2019, DLAP 2019 in Kyoto and Corfu workshop on {\it Recent Developments in Strings and Gravity 2019.}
LA and JG are supported in part by NSF grant PHY-2014086. NR is supported by NSF grant PHY-1720321. 

\appendix
\section{Sampling}
\label{app:Sampling}

This appendix summarizes known results from the literature about sampling and summarizes our conventions. We start by discussing two methods for sampling points on the hypersurface. We then present a simple example on why restricting to the CY hypersurface can lead to a point sample with a non-flat prior and discuss how our re-weighting of points is implemented.

\subsection{Sampling by solving for the dependent coordinate}
We use the following method to generate training data for our metric neural networks utilizing one network per coordinate patch, which are presented in Section~\ref{sec:learningthemetric}.

Recall that to obtain an affine patch of an $n$-dimensional variety $X$, we go to a patch $U_i$ of the ambient space $\bP^{n+1}$ where $z_i=1$, and we solve for a coordinate $z_j$ with $j\neq i$. Thus, a basic approach for generating points on $X$ is to first sample $n$ complex numbers $\qty(z^{(i)}_0, \ldots, z^{(i)}_{n+1})$, where we have skipped over $z^{(i)}_i$ and $z^{(i)}_j$. We then solve $p_\psi\qty(\vec{z}^{(i)}) = 0$ for $z^{(i)}_{j}$ to obtain affine coordinates in patch $U_i$ of the ambient space $\bP^{n+1}$.
These numbers, along with information specifying the chart $U_i$ of the ambient space, uniquely define a point on $X$.
Depending on the manifold, one may have to restrict the sampling of the initial coordinates so that the equation $p_\psi\qty(\vec{z}^{(i)}) = 0$ has a solution for the last coordinate. Note that, if the defining polynomial is symmetric under coordinate permutation, one may be able to use the coordinates generated for a point on one patch to immediately obtain points on other patches.

For example, in the case of the Fermat quintic ($\psi=0$ and $n=3$), we have in the affine patch $U_0$ that
\begin{align}
  p_0\qty(\vec{z}^{(0)}) = 1 + \sum_{i=1}^{4} \qty(z^{(0)}_i)^5 \,.
\end{align}
If $j$ is $4$, one can then solve for $z_4^{(0)}$ given $\qty(z_1^{(0)}, z_2^{(0)}, z_3^{(0)})$:
\begin{align}
  z_4^{(0)} = \sqrt[5]{-1-\sum_{i=1}^{3} \qty(z_i^{(0)})^5} \,.
\end{align}
 Since there are in general five fifth roots, one gets for each choice of initial complex values five points on $X$.
 
The crucial step is to find the solutions of the single-variable complex polynomial equation $p_{\psi}(\vec{z})=0$ (with all but one affine coordinate fixed).
A fast method to do this is by computing the eigenvalues of the polynomial's companion matrix.

\subsection{Illustration of rejection sampling}
\begin{figure}[t]
\centering
\includegraphics[width=0.4\textwidth]{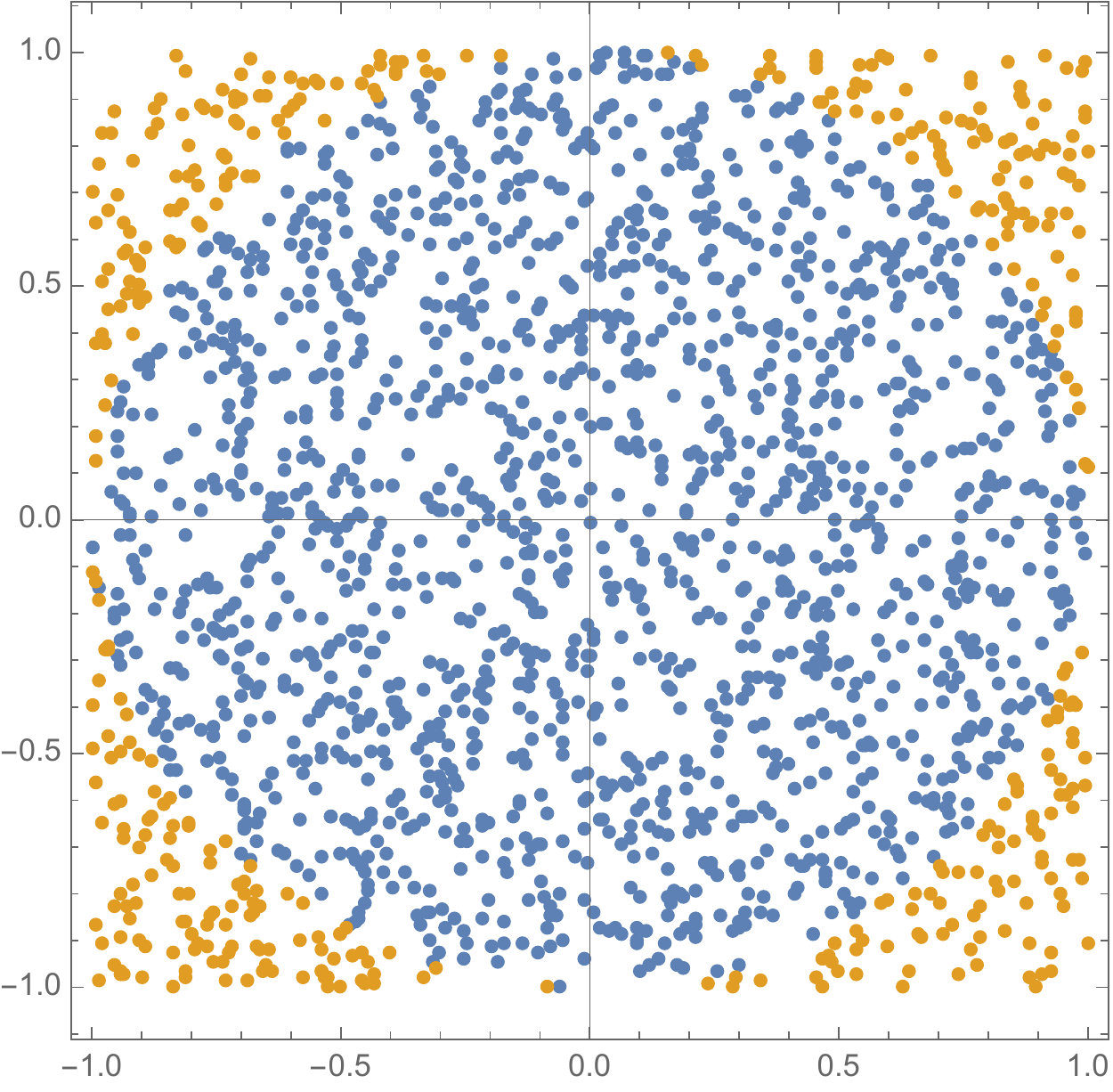}\quad
\includegraphics[width=0.4\textwidth]{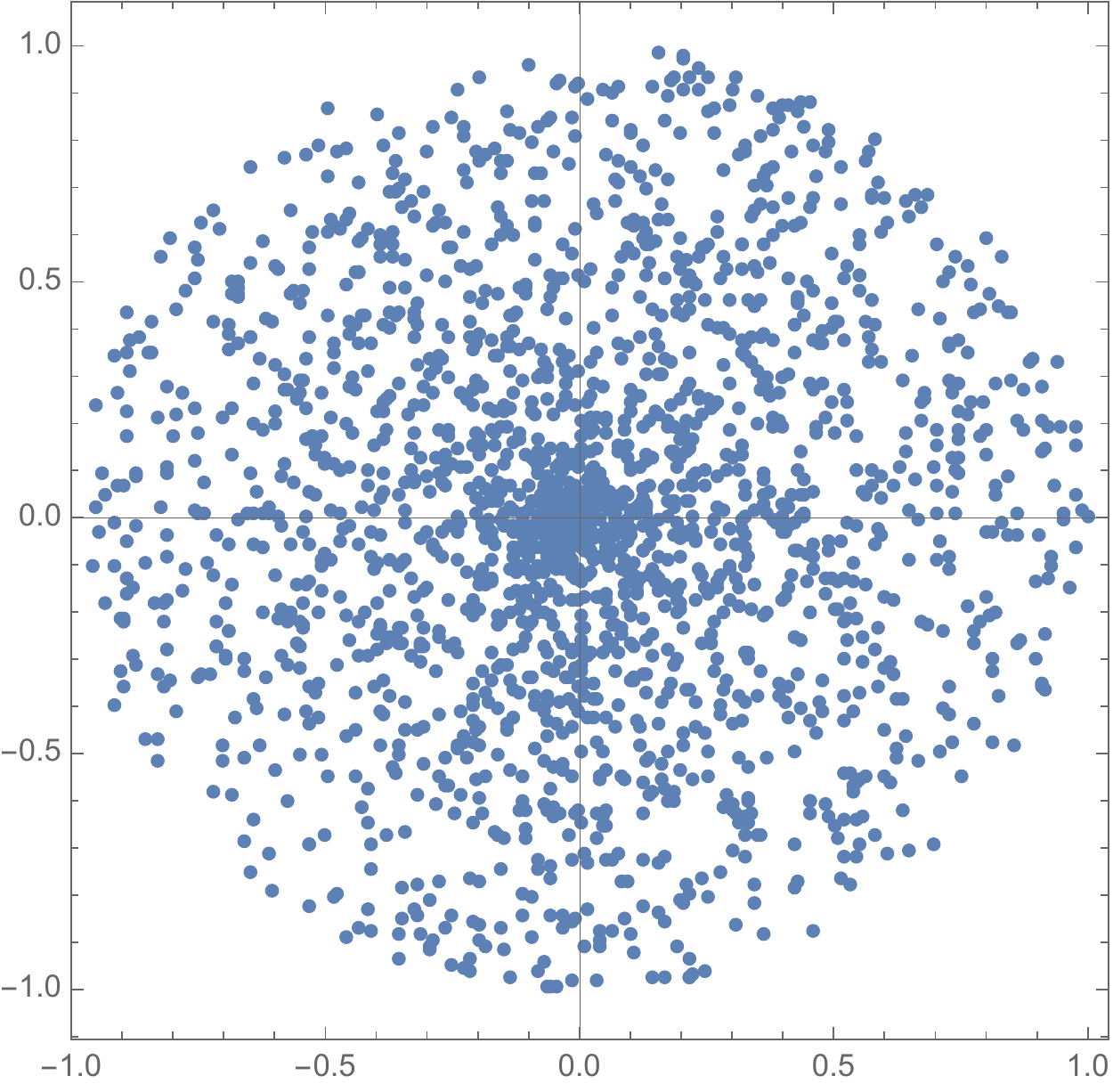}
\caption{Illustration of rejection sampling (left) and how the induced measure for sampling spherical coordinates with a flat prior can be non-flat (right) using the example of the unit disk.}
\label{fig:samplingPoints}
\end{figure}

To illustrate that the measure when restricting to the CY hypersurface is non-flat, let us consider points on the unit disk. One way of getting a flat distribution of points inside the disk would be to just randomly sample points in the interval $[-1,1]\times[-1,1]$ and throw away those points that do not lie inside the disk, cf.\ the left-hand-side of Figure~\ref{fig:samplingPoints}. This type of rejection sampling works in our case as well, but it is extremely ineffective, especially at larger $\psi$. So if one used spherical coordinates $x+iy=re^{i\varphi}$ and sampled with a flat prior $r\in[0,1]$, $\varphi\in[0,2\pi]$, one would get only points inside the unit disk. However, as shown in Figure~\ref{fig:samplingPoints} on the right, the induced measure on the disk is not flat. In our case, the way to correct the auxiliary measure to account for this sample bias (i.e.~how to compute the weights of each point) is explained in~\cite{Douglas:2006rr}, and we comment below on the implementation.

\subsection{Homogeneous sampling in projective space}

For all other network discussed in this paper, we sample points on the manifold $X$ using intersections with a line, which involves the following steps:
\begin{enumerate}
  \item Uniformly sample two points $\vec{a}, \vec{b} \in \bP^{n+1}$, thereby defining a complex line.
  \item Compute the following polynomial in the complex variable $t$:
    \begin{align}
      p_\psi(\vec{a} + t \vec{b}) = 0 \,,
    \end{align}
    where $p_\psi(\vec{z})$ is the defining homogeneous polynomial of $X$.
    This can either be done manually given a specific defining equation, or using a library for symbolic manipulations. This was done for the implementation here using SymPy~\cite{SymPy}, making it more easily extendable to other defining equations)
  \item Solve the defining equation for $t$, for example by finding the eigenvalues of the polynomial's companion matrix or simply numerically.
  \item Due to the multiplicity of roots, each chosen line intersects the manifold in $n+2$ points $\vec{z} = \vec{a} + t \vec{b}$. 
\end{enumerate}

One can uniformly sample points on $\bP^{n+1}$ by first sampling real numbers from $S^{2(n+2)}$ and combining them into complex numbers representing homogeneous coordinates.
There are multiple algorithms for sampling points on a real sphere; an efficient one is to independently sample coordinates from a normal distribution and then divide by their norm.

Since the line is chosen uniformly in projective space, this sampling algorithm leads to points on the manifold that are not uniform to its volume form, but uniform with respect to the Fubini-Study metric on the ambient space.

\begin{figure}[t]
  \centering
  \includegraphics[width=0.9\textwidth]{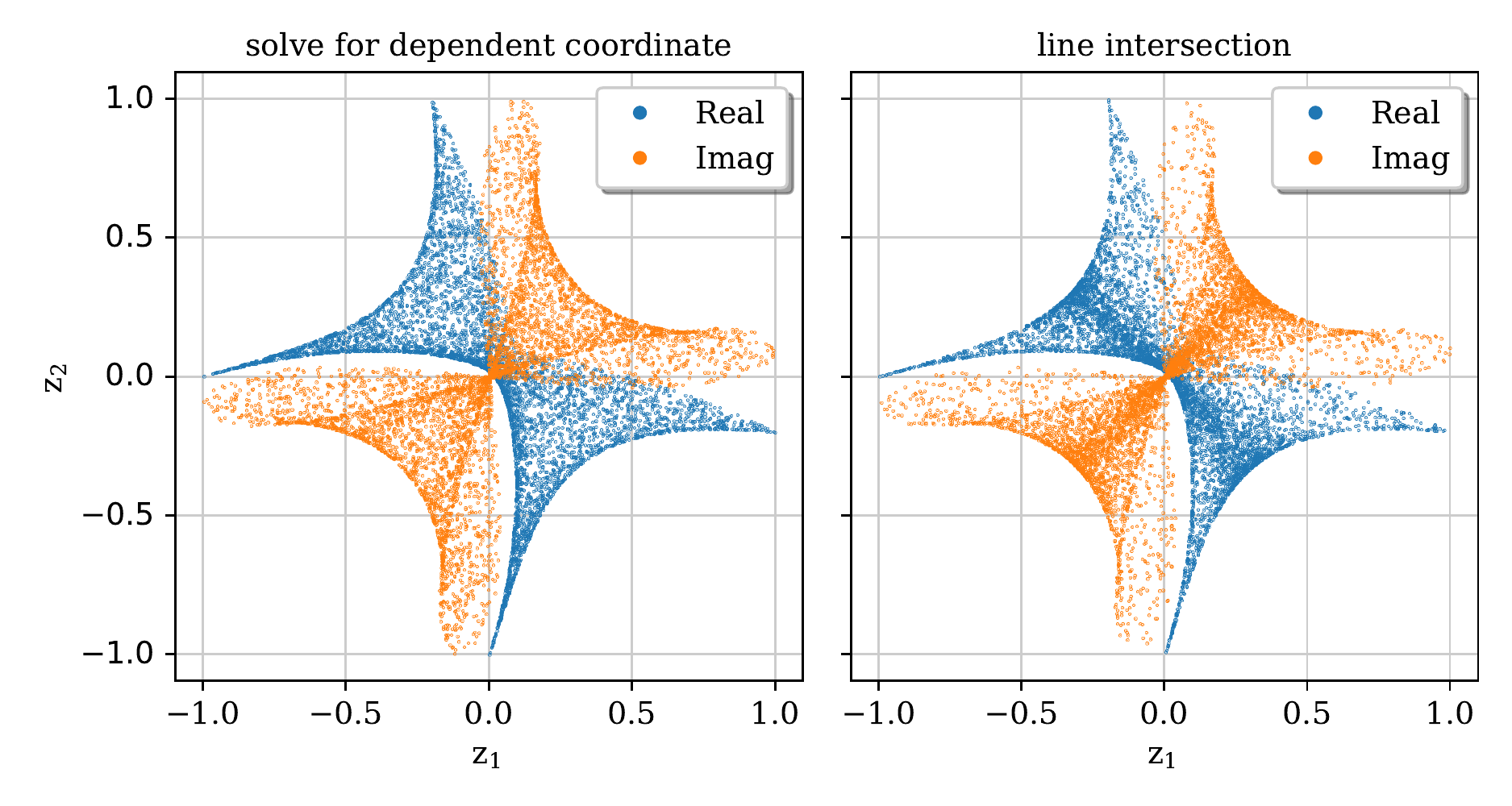}
  \vspace{-15pt}
  \caption{Scatter-plot comparing real and imaginary parts of the affine coordinates of the variety in $\CP{2}$ defined by ${z_0^3 + z_1^3 + z_2^3 + 10 \, z_0z_1z_2=0}$, generated using either sampling algorithm introduced in Section~\ref{app:Sampling}.
  The values all lie in the affine patch $\hat{U}_0$ defined by $|z_1|,|z_2| \leq |z_0| = 1$.}
  \label{fig:sampling-example}
\end{figure}

A specific example of the difference between the two sampling algorithms defined above can be found in \figref{fig:sampling-example}.

\section{Algebraic metrics and Donaldson's algorithm}

\subsection{Constructing the monomial basis}
\label{app:Polynomials}
When the basis of the line bundle $\mathcal{O}_{\CP{n+1}}(k)$, given by all homogeneous monomials defined in the homogeneous projective coordinates, is restricted to $X$, the basis has to be reduced for $k \geq n+2$.
The reason is that on $X$ the defining polynomial , $p_\psi$ vanishes, which means that all polynomials containing $p_\psi$ (a degree $n+2$ polynomial) must be removed to obtain a basis.
Formally, the basis is defined as
\begin{align}
  \Comp[z_0, \ldots z_{n+1}]_k \,/ \left<p_{\psi}(\vec{z})\right> \,,
\end{align}
where
\begin{align}
  \left<p_\psi(\vec{z})\right> = p_\psi(\vec{z}) \, \Comp[z^0, \ldots z^{n+1}]_{k-(n+2)} \,.
\end{align}
Another perspective on this is that each linearly independent polynomial in $\left<p_\psi(\vec{z})\right>$ can be rewritten to express one of the constituent monomials in terms of the remaining monomials.
We get the following expression for the number of basis sections of $\mathcal{O}_X(k)$:
\begin{align}
\label{eq:numberSections}
N_k=\left(\begin{array}{c}k+4\\k\end{array}\right)-\left(\begin{array}{c}k-1\\k-5\end{array}\right)\,.
\end{align}
The second term is precisely the number of sections that become linearly dependent under pullback. (We follow the convention that a binomial coefficient with negative entries is zero).

To make this clearer, consider $k=6$, $n=3$, and $p_\psi(\vec{z}) =  \sum_i z_i^5 + \psi \prod_i z_i$.
A basis of $\langle p_\psi(\vec{z}) \rangle$ is then given by multiplying $p_\psi$ with the basis $\qty{z_0, z_1, z_2, z_3, z_4}$ of $\Comp[z_0, \ldots, z_4]_1$.
Since $p_\psi$ vanishes, the following relations are generated
\begin{align}
  z_j \, \qty(\sum_i z_i^5 + \psi \prod_i z_i) = 0 \quad \forall j \,.
\end{align}
Each of these $5$ equations can be used to eliminate one monomial.
One choice is to remove all monomials $z_j \, z_0^5$, $j=0, \ldots, 4$.

So far, the discussion of how the reduced monomial basis is obtained was on a mathematical level.
In practice, the sections can be represented using a matrix of integers.
For example, the monomial
\begin{align}
  s(\vec{z}) = z_0 \, z_2^3 \, z_3
\end{align}
of $\mathcal{O}_{\CP{3}}(5)$ corresponds to the row vector
\begin{align}
  \qty[1, 0, 3, 1] \,.
\end{align}
The full basis of monomials of, for example, $\mathcal{O}_{\CP{1}}(2)$ can be written as a matrix with each row representing a monomial:
\begin{align}
  \mqty[2 & 0 \\ 1 & 1 \\ 0 & 2 ] \,.
\end{align}
Given a defining equation such as $p(\vec{z}) = z_0^2 + z_1^2$, each summand corresponds to a row in the matrix.
Solving for either summand and removing it from the basis thus corresponds to deleting a row in the matrix.
For the current example, either of $[0, 2]$ and $[2, 0]$ could be removed to obtain a basis on $X$.
Both the generation of the monomial basis on projective space, and the reduction given a defining polynomial can be done algorithmically.
This allows the defining equation to be replaced without adding significant implementation work.

\subsection{Donaldson's algorithm}
\label{sec:Donaldson}
Extending work of Tian~\cite{Tian:1990aaa}, Donaldson presented in~\cite{Donaldson:2005mat} an approximation scheme for Ricci-flat CY metrics, which lends itself to numerical implementation on a computer. Indeed, the method was adopted in the physics literature soon afterwards~\cite{Douglas:2006rr,Braun:2007sn,Ashmore:2019wzb}. The algorithm relies on the CY manifold $X$ having an embedding into projective spaces (whose homogeneous coordinates we denote collectively by $\vec{z}$) and uses numerical integration paired with an iteration procedure to approximate the Ricci-flat metric.

The algorithm is described in detail in~\cite{Douglas:2006rr,Braun:2007sn,Ashmore:2019wzb}, so we will just outline the different steps. We have implemented the algorithm in Mathematica and JAX~\cite{jax2018github}. To test our implementations, we compared the results with~\cite{Braun:2007sn,Ashmore:2019wzb}. The algorithm finds the balanced metric as follows:

\begin{enumerate}
\item Choose a (multi-) degree $k$ of an ample line bundle to work with. The approximation error was proven to go to zero as $k\to\infty$. The (multi-) degrees fix a direction in the Picard lattice dual to the K\"ahler cone. 
\item Find a basis of sections $s_\alpha$, $\alpha=1,\ldots, N_k$ of the line bundle which restrict non-trivially to the CY manifold $X$ in question (where $N_k$ depends on $k$). This is described in Section~\ref{app:Polynomials}.
\item Fix a complex structure and find points $\vec{z}_i$, $i=1,\ldots,N_p$ on $X$ for this choice (e.g.\ by intersecting a line defined by two randomly, uniformly distributed points in the ambient space with the CY manifold). See Appendix~\ref{app:Sampling} for the implementation.
\label{enum:ComputePoints}
\item Compute the weights $w_i$, $i=1,\ldots,N_p$ of the induced distribution of sampled points on $X$. (These are not drawn from a flat prior even though the ambient points were.)  In terms of these weights, the numerical integration reduces to
\begin{align}
  \int_X \Omega\wedge\bar\Omega \to \frac{1}{N_p}\sum_i w_i\,,\qquad \int_X J^3 \to \frac{1}{N_p}\sum_i w_i\left.\frac{J^3}{\Omega\wedge\bar\Omega}\right|_{\vec{z}=\vec{z}_i}\,.
\end{align}
\item Choose a random initial Hermitian $N_k\times N_k$ matrix $H^{(0)}_{\alpha\bar{\beta}}.$
\item Compute %
\begin{align}
  \tilde{H}_{\alpha\bar{\beta}}^{(\ell)} = \frac{N_k}{\sum w_i} \;\frac{\sum_i w_i\, s_\alpha(\vec{z}_i) \bar{s}_{\bar{\beta}}(\vec{z}_i)}{\sum_i s_\gamma(\vec{z}_i) H^{(\ell)}_{\gamma\bar{\delta}} \bar{s}_{\bar{\delta}}(\vec{z}_i)}~.
\end{align}
The sum over the points and the weights appear from the numerical integration.
\item Set $H^{(\ell+1)}=\qty(\tilde{H}^{(\ell)})^{-1}$ and return to the previous step. Alternatively, sample new points and re-calculate the weights and then go to step 6.
\item Repeat until we reach a fixpoint, i.e.~$H^{(\ell+1)} \approx H^{(\ell)}$. In practice around 10-20 steps are typically enough. We terminate the procedure either after a certain number of steps or when the maximum absolute value of the difference of  $H^{(\ell+1)}$ and  $H^{(\ell)}$ is smaller than $10^{-6}.$
\item The Ricci-flat K\"ahler metric is given in terms of the K\"ahler potential %
\begin{align}
  K=\frac{1}{2\pi k}\ln(s_\alpha H_{\alpha\bar{\beta}} s_{\bar{\beta}})
\end{align}
From this example, we see that for $k=1$, $s=\vec{z}$, and $H=\mathbbm{1}_{(d+2)\times(d+2)}$, this is just the FS K\"ahler potential.
\end{enumerate}
The metric found in this fixpoint procedure is called balanced.

In order to arrive at an expression for the CY metric $g_\CY$, we need to perform two more steps. First, we need to account for the projective rescaling degrees of freedom. This is best done by going to an affine patch. We go to the patch where we scale the coordinates with the largest absolute values to unity in order to ensure numerical stability of the algorithm. We denote the affine patch coordinates by $\vec{z}$. 

Second, we need to pull back the metric computed from the K\"ahler potential, which is produced by the algorithm above, to the CY manifold. On the CY space $X$, we can think of $m$ of the remaining $m+3$  affine coordinates as being (implicit) functions of the others. Since the $(3+m)\times(3+m)$ metric $\hat{g}$ in an affine patch but prior to pullback is given by
\begin{align}
\hat{g}_{a b} =\partial_a\overline{\partial}_b K\,,
\end{align}
the $3\times (3+m)$ pullback map is given by
\begin{align}
C_{\mu a}=\frac{\partial z_a}{\partial x_\mu}~,
\end{align}
where the $x_\mu$ are local coordinates on $X$. It should be noted that this can be computed in terms of derivatives of the defining equations with no need to actually solve the equations for the $m$ coordinates that are to be eliminated. The pulled back metric is then
\begin{align}
g_\CY = i^*(\hat{g})= C\cdot \hat{g}_\CY \cdot C^\dagger~.
\end{align}

\subsubsection*{Implementation of Donaldson's algorithm in pseudocode}
Below is a simplified Python pseudocode which illustrates how a single iteration of Donaldson's algorithm is computed.

\begin{lstlisting}[language=Python, caption={Simplified algorithm for computing a single iteration of Donaldson's algorithm.}, label={alg:donaldson}, abovecaptionskip=5pt, mathescape=true]
def donaldson_step(variety, H, k, pows, $\psi$, vol_cy, count):
  # Approximate the T operator for the `H`-matrix of degree `k`
  # by a Monte Carlo sum over `count` sample points.

  # accumulate the integral in this variable
  T = zeros_like(H)

  for i in range(count):
    # pretend this returns a single point now, for simplicity
    z = line_sample(variety, $\psi$)
    z, patch = to_affine(z)

    weight = mc_weight(variety, z, patch, $\psi$)
    $s$ = compute_monomials(z, patch, k)
    $\bar{s}$ = conj($s$)

    numerator = $s_\alpha \bar{s}_{\bar{\beta}}$
    denominator = $s_\alpha H_{\alpha\bar{\beta}} s_{\bar{\beta}}$

    dT = numerator / denominator * weight
    T = T + dT / count

  T = T * basis_size(variety, k) / vol_cy
  new_H = invert(T).transpose()
  return new_H
\end{lstlisting}

\subsubsection{Finding equivariant elements via Donaldson's algorithm}
\label{app:donaldsonelements}
\begin{figure}
\includegraphics[width=1\textwidth]{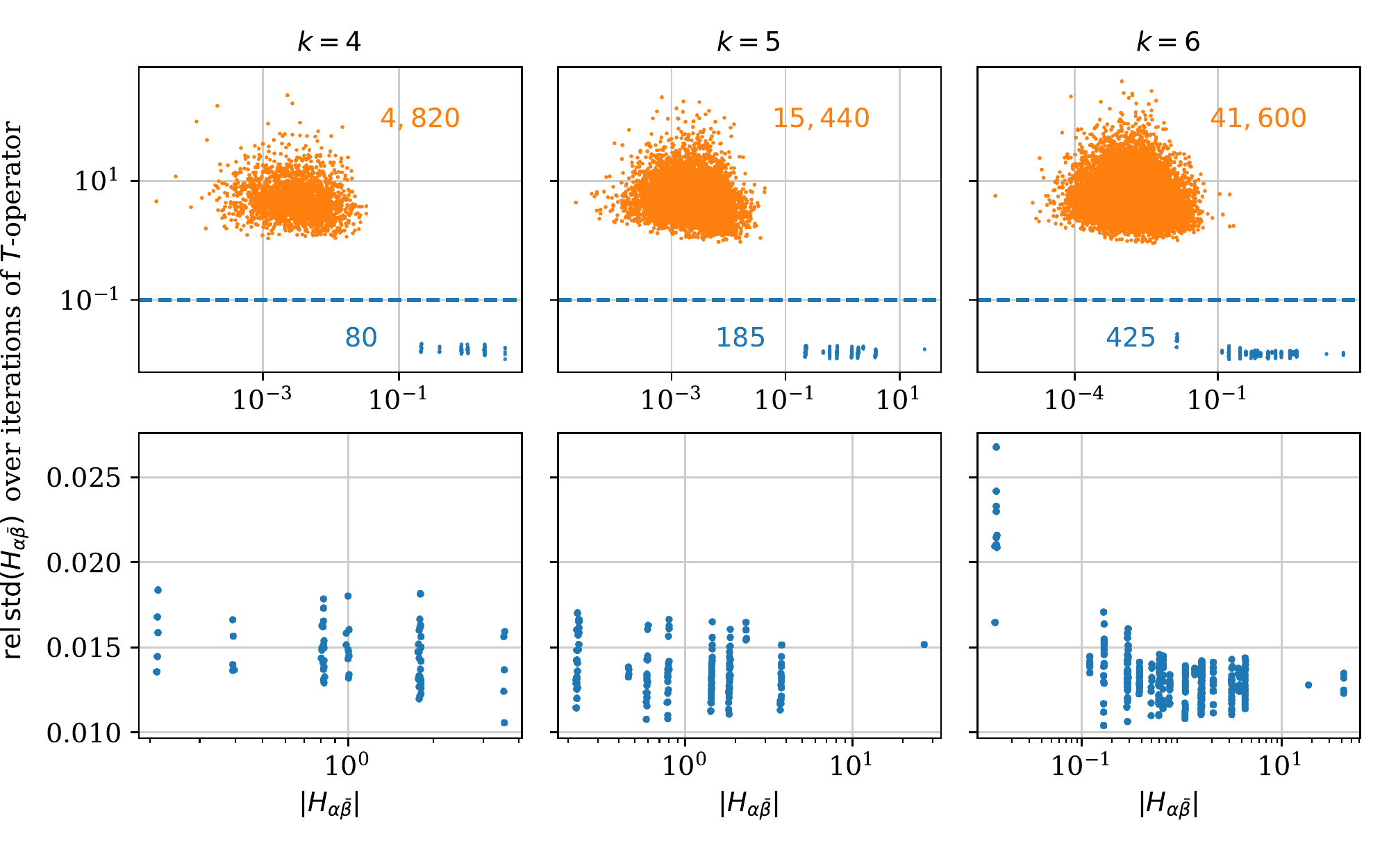}
  \caption{Clustering of elements in $H$ for $\psi=10$. {\bf Top row:} Orange clusters correspond to vanishing elements, blue clusters correspond to non-vanishing values. {\bf Bottom rows:} Close-up view of the blue clusters. The number of vertical lines is in close relation to the number of equivariant components.}\label{fig:donaldsonpattern}
\end{figure}

When evaluating the relative standard deviation over iterations of Donaldson's algorithm for multiple values of $k$ and $\psi$ in relation to their absolute value, we can identify two clusters as shown in Figure~\ref{fig:donaldsonpattern}. The blue cluster, containing most of the components, corresponds to elements which are essentially vanishing and the fluctuations are relatively large. The orange cluster has small fluctuations but includes several elements which are small. The number of vertical lines in the orange cluster matches with the number of invariant polynomials under the discrete symmetries as it should. This provides a cross-check that the numerical approximation is valid. Conversely, it can serve to detect underlying symmetries in the K\"ahler potential.\footnote{See~\cite{Krippendorf:2020gny} for work on detecting symmetries in theoretical high-energy settings using NNs.} We leave a more systematic study of this observation for the future.

\section{Details for training \textit{H} networks}
\label{app:hdirectly_details}
In this Appendix, we provide more details on the experiments we have performed for learning the $H$ matrix with and  without using data obtained using Donaldson's algorithm.

\subsection{Supervised training with Donaldson's algorithm}
\label{app:h_supervised}

In designing and training the NN, we found that the result is not very sensitive to hyperparameter tuning and does not require complicated network architectures. For this paper, we chose a simple feed-forward NN with 3 hidden layers of dimensions 100, 2000 and 2000 with (leaky) ReLU activation, cf.\ Table~\ref{tab:NN-H}. The input is (the real part, imaginary part, and absolute value of) $\psi$ and the output are the $N_k^2$ independent (real and imaginary) components of $H$.\footnote{We ran experiments where we added (the real and imaginary part of) powers of $\psi$ to the input. However, for large $\psi$, positive powers tend to produce rather large features. So one should either normalize them to unit variance (since we draw $\psi$ randomly from a flat prior, it will already have roughly zero mean), which is problematic if one wants to extrapolate beyond the training set. For fractional powers, one will have to choose a branch or include all branches as features. Since the observed accuracy improvements are rather small, we ended up using $\text{Re}(\psi)$, $\text{Im}(\psi)$, and $|\psi|$ as features.}

\begin{table}[t]
\centering
\begin{tabular}{|c|c|c|c|}
\hline
Layer & Number of Nodes & Activation & Number of  Parameters\\
\hline
input & 3 & -- & --\\
hidden 1 & 100 & leaky ReLU & 400\\
hidden 2 & 1000 & leaky ReLU & 101\,000\\
hidden 3 & 1000 & leaky ReLU & 1\,001\,000\\
output & $N_k^2$ & identity & 1000$ \times N_k^2+N_k^2$\\
\hline
\end{tabular}
\caption{Neural network architecture for the neural network that learns the $\psi$-dependence of $H$.}
\label{tab:NN-H}
\end{table}

As explained above, we choose the patch where we set the largest absolute value of the coordinates to 1 and solve implicitly for the coordinate for which the derivative of $p$ has the largest absolute value. With these results, we compute $\sigma$ as defined in~\eqref{eq:sigma}, which is between $0.14$ and $0.39$ for the quintic with $k=3$ and $\psi$ in the specified range.\footnote{We observe that $\sigma$ gets larger as $|\psi|$ gets larger.} Hence, even if the NN computing $H$ had zero error, the numerical error dictated by using $k=2$ would be $0.2$ when using $\sigma$ as a measure for precision. In our experiment, we trained the network with 90 percent of the grid points and evaluated on the remaining 10 percent. We train the NN for 200 epochs with stochastic gradient descent, ADAM optimizer and L$_2$ weight decay with parameter $0.001$. This takes less than a minute and is orders of magnitude faster than re-computing $H$ for a given value of $\psi$.

\subsection[Learning \textit{H} by minimizing the Monge-Amp\`{e}re loss (at constant \texorpdfstring{$\psi$}{psi})]{Learning \textit{H} by minimizing the Monge-Amp\`ere loss (at constant \texorpdfstring{$\boldsymbol\psi$}{psi})}
Before training $\psi$-dependent networks that output $H$, consider the case of fixed $\psi$ and $k$.
We now want to find the optimal matrix $H$, defined as the one that minimizes the Monge-Amp\`ere loss at the given degree $k$.
This is precisely the situation that was explored in~\cite{Headrick:2009jz}.
As a first step towards ML, we have repeated the optimization using stochastic gradient descent. 
The main difference is that instead of picking a large set of points on the manifold and finding $H$ by least-squares, we use multiple steps of gradient descent, each time computed over a random batch of fewer points.
Choosing a different random sample of points for each batch has the advantage that the number of points used can be decreased, while avoiding over-fitting.
We have replicated the results in~\cite{Headrick:2009jz} for degrees up degree $k=6$ and several values of $\psi$.
This establishes the basic stochastic gradient descent setup that will be used for the more complex models.

\subsection{Learning \textit{H} by minimizing the Ricci loss}
\label{sec:ricciloss}

The second type of loss introduced in Section~\ref{sec:Accuracy} is one based on minimizing the Ricci curvature. Here we use the Ricci scalar as a loss function
\begin{align}
  R_0 = \frac{1}{M} \sum_{a=1}^M w(z_a) \, |R(z_a)|^2 \,,
  \label{eq:ricci-loss}
\end{align}
where $M$ is the number of points and $w(z_a)$ is the associated weight.
Because this loss depends on the K\"ahler potential in its fourth derivative, it is significantly more expensive to compute than the Monge-Amp\`ere loss applied in the rest of the paper.
Where the $H$ matrix converged within minutes for the Monge-Amp\`ere loss with $k \leq 6$, the Ricci-based loss converged within tens of minutes.

\begin{figure}[t]
  \centering
  \includegraphics[width=1\textwidth]{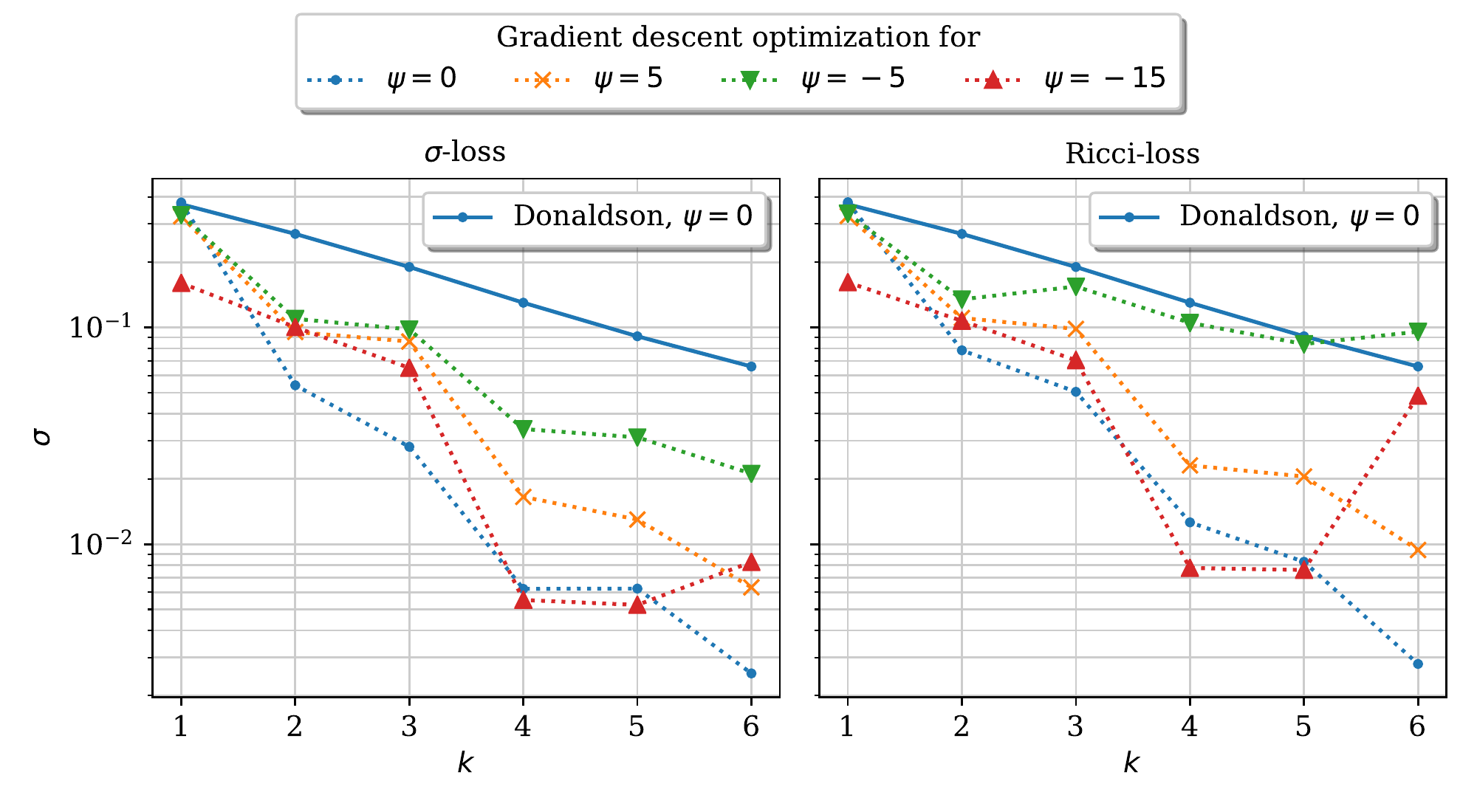}
  \vspace{-12pt}
  \caption{$\sigma$-accuracies achieved by an optimization of the $H$ matrix using gradient descent with respect to the Monge-Amp\`ere loss on the left, and the Ricci loss of Equation~\eqref{eq:ricci-loss} on the right.}
  \label{fig:gradient}
\end{figure}

\figref{fig:gradient} shows  the $\sigma$ accuracies achieved by a similar gradient descent setup as in the previous section, using instead the Ricci-scalar loss defined in Equation~\eqref{eq:ricci-loss}.
Both losses should have the same global minimum with respect to the $\sigma$ measure at each degree $k$. No exhaustive parameter search was conducted to obtain optimal convergence in either case, so the results should be understood as showing both losses are feasible and lead to approximations of the flat metric. Convergence is similar to the one achieved using the Monge-Amp\`ere loss.
This shows that gradient descent is in principle also possible for more complicated loss functions, depending on higher derivatives of the Kähler potential.
Due to its higher complexity, we have not pursued the Ricci loss further in this work. However, it has the advantage that it can be extended to the case of constant Ricci scalar, which is not further investigated here:
\begin{align}
  R_c = \frac{1}{M} \sum_{a=1}^M w(z_a) \, |R(z_a) - c|^2 \,,
\end{align}
where $c$ denotes the target curvature.

\subsection[\textit{H} networks with \texorpdfstring{$\psi$}{psi} dependence]{\textit{H} networks with \texorpdfstring{$\boldsymbol\psi$}{psi} dependence}

We now want to find a network that describes a map from $\psi$ to the Hermitian matrix $H$.
Since within the network we want to work with real numbers, we first have to choose how to map the complex value $\psi$ to a set of real input features. We have tried several possibilities:
\begin{itemize}
  \item Split into $|\psi|$ and $\text{arg}(\psi)$.
  \item Introduce an additional array of powers and compute $|\psi|^{p_i}$.
  \item Raise to a power and split into real and imaginary parts, $\text{Re}[\psi^{p_i}]$, $\text{Im}[\psi^{p_i}]$.
\end{itemize}

Following the choice of input features, we add some number of dense layers (how many are best seems to depend on the range of $\psi$ that we want to optimize over), each with a sigmoid activation function.
This number of dense layers is referred to above as the number of hidden layers.
In order to get the right number of parameters, a final dense layer with trivial activation function maps from the last features to the required number of values.

During our experiments we observed that it is beneficial to multiply the final output parameters by a modulation factor as in
\begin{equation}
 H_{\rm out}=\sigma(\tilde{H})\hat{H}~,
\end{equation}
where $\sigma$ denotes the sigmoid-function. This makes it more stable for gradient descent to set some output values to zero. Besides parametrizing the real and imaginary parameters of the Hermitian $H$ matrix directly, we also used a parametrization via the following Cholesky decomposition:
\begin{gather}
  L = \mqty(
    H^{\text{d}}_{1} && H^{\text{r}} + i H^{\text{i}} \\
    & \ddots & \\
    \phantom{H^r}0\phantom{iH^i} && H^{\text{d}}_{N_k} \\
  ) \,, \quad
  H = L L^\dag \,,
\end{gather}
where now the diagonal entries are positive.
This prevents negative or zero eigenvalues, which may lead to non-definite metrics on the manifold.
Our experiments indicate that this leads to slightly more stable gradient descent training.

\begin{figure}
  \includegraphics[width=1\textwidth]{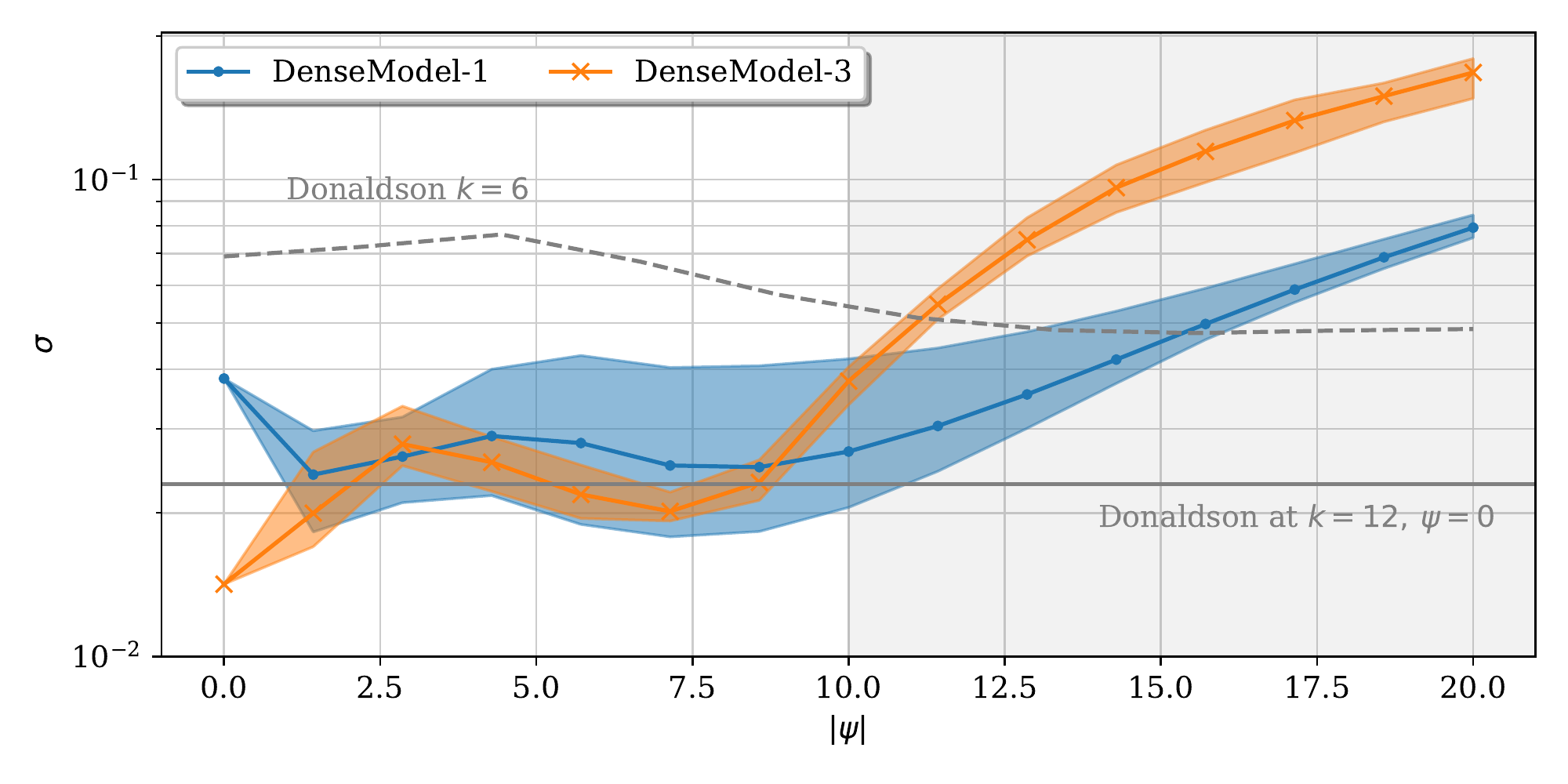}
  \caption{Accuracies achieved by two network architectures detailed in Section~\ref{app:densenet} trained over uniform values $|\psi|<10$ using the MA loss. The shaded area points out the extrapolation of the networks. As a comparison, the $\sigma$ accuracies achieved by Donaldson's algorithm over real values of $\psi$ are shown. The error band indicates the minimal and maximal values measured over multiple complex angles.}
  \label{fig:hpsi-smallscale}
\end{figure}

\subsection{Network architectures}
\label{app:densenet}
The following is a brief summary of different network architectures used to produce the results in the above figures. 
\subsubsection*{DenseModel-1}
For the first model, we start with the input features $|\psi|$ and $\text{arg}(\psi)$, followed by a single hidden layer of dimension equal to the basis size $N_k$, in our case $N_6=205$. This is motivated by the fact that due to our choice of symmetric manifold, we expect relatively few independent components of $H$. To construct the final $H$ matrix, we use the Cholesky decomposition.

\subsubsection*{DenseModel-2}
This model is exactly the same as the above, except that now we use two hidden layers of dimension $N_k$, each.

\subsubsection*{DenseModel-3}
As input features we take the real and imaginary parts of $\psi$ raised to the powers $1, 2, 3,$ and $1/2$. This is followed by a single hidden layer of size $N_k$. The output $H$ is constructed using the decomposition of a Hermitian matrix into real and imaginary components.

\section{Details on metric training}
\label{sec:hyperparametermetric}
Here we provide more details about the training of the metric learning networks and our hyperparameter choices.  

\subsection[NN with homogeneous coordinate input (\texorpdfstring{$\psi=10$}{psi=10})]{NN with homogeneous coordinate input (\texorpdfstring{$\boldsymbol{\psi=10}$}{psi=10})}
For this input and output setup, we have performed some hyperparameter searches. We tried learning rates of $10^{-i}$ with $i\in[3,6]$ and $L_2$ weight decays with parameter $10^{-i}$ with $i\in[4,8]$. We also tried different activation functions (leaky ReLU, GELU, ELU, Tanh) and optimizers (ADAM, Adagrad, SGD) as well as varying the number of hidden layers and the nodes in the hidden layers. We also included dropout or batch norm layers, but this did not significantly change our results. In the end, we got good results already for a rather small, simple, feedforward neural net (without dropout or batch norm), with learning rate $10^{-4}$, no weight decay, leaky ReLU activation and ADAM optimizer. We chose a rather large batch size of 900 (memory-wise, this is not a problem since each individual training sample is not too big). We summarize the architecture and the number of parameters in Table~\ref{tab:NN-g}. As explained in Section~\ref{sec:learningthemetric}, the input to the neural network consists of the real and imaginary part of the point on the quintic expressed in homogeneous ambient space components (10 nodes), of the real and imaginary part of $\psi$ (two components), and of a True/False encoding of which of the ambient space coordinates is used for pulling back and as a patch coordinate (5 components). For the sake of concreteness, we compute and compare the CY metrics at $\psi=10$ on a dataset with 50000 points, which we split according to train:test=90:10, and we train for 20 epochs. During training, we monitor the training and the test loss and stop earlier if they start to diverge (which does not happen).

We found that the linear metric perturbation only improves the error measure $\sigma$ marginally as compared to the FS metric. We also observed that the overlap and K\"ahler loss grow rapidly for the additive ansatz if we do not actively optimize for them in contrast to the multiplicative ansatz. This means that the parameters $\lambda_i$ in~\eqref{eq:totalLoss}  need to be rather fine-tuned in the former case. For these reasons we focus on the multiplicative loss. The results were shown in the main text (cf.~Section~\ref{sec:resultsmetric}) for $\psi=10$.

\begin{table}[t]
\centering
\begin{tabular}{|c|c|c|c|}
\hline
Layer & Number of Nodes & Activation & Number of Parameters\\
\hline
input & 17 & -- & --\\
hidden 1 & 100 & leaky ReLU & 1800\\
hidden 2 & 100 & leaky ReLU & 10\,100\\
hidden 3 & 100 & leaky ReLU & 10\,100\\
output & $d^2$ & identity & 101 $d^2$\\
\hline
\end{tabular}
\caption{Neural network architecture for the neural network that approximates the CY metric directly by optimizing a loss function that combines the various consistency conditions for the CY metrics.}
\label{tab:NN-g}
\end{table}

Finally, we want to remark that these results do not change if we include $\psi$ as an input to the NN and train it for different complex structures. A different approach would be to learn the metric $g$ for different (fixed) $\psi$ and train a second NN that interpolates $g$ (instead of $H$ as described in Section~\ref{sec:supervised_donaldson}). Since providing $\psi$ as additional input to the NN in the training process worked well, we have not pursued this second option further.

\subsection[NN with affine coordinate input (LDL output \texorpdfstring{${0<|\psi|<10}$}{0<psi<10})]{NN with affine coordinate input (LDL output \texorpdfstring{$\boldsymbol{0<|\psi|<10}$}{0<psi<10})}
This class of NNs has one NN for each patch and the output is in the LDL decomposition of the metric~\eqref{eq:ldldecomposition}. We then readily compute the metric from this output. Each of these networks has trainable parameters as shown in Table~\ref{tab:NN-g2}. The initialization is chosen such that the initial network is close to the Fubini study metric. During training we monitor how close the network is to the Fubini Study metric.

\begin{table}[t]
\centering
\begin{tabular}{|c|c|c|c|c|}
\hline
Layer & number nodes & Activation & Regularization & Initialization \\
\hline
input & 10 & -- & --& --\\
hidden 1 & 1000 & ReLU & L2($10^{-6}$) & ${\cal N}_k(0,10^{-4})$, ${\cal N}_k(0,10^{-3})$\\
hidden 2 & 1000 & ReLU & L2($10^{-6}$) & ${\cal N}_k(0,10^{-4})$, ${\cal N}_k(0,10^{-3})$\\
hidden 3 & 1000 & ReLU & L2($10^{-6}$) & ${\cal N}_k(0,10^{-4})$, ${\cal N}_k(0,10^{-3})$\\
output & 9 & -- & L2($10^{-4}$) & ${\cal N}_{k,b}(0,10^{-2})$\\
\hline
\end{tabular}
\caption{Neural network architecture for each `patch' neural network that approximates the CY metric.}
\label{tab:NN-g2}
\end{table}
We have trained this network using $50000$ points for each patch and overlap region and validated the network using $10000$ points respectively in the range $0<|\psi|<10.$ The respective loss weights were $\lambda_{MA}=1,$ $\lambda_{\rm overlap}=0.1,$ and $\lambda_{dJ}=0.1.$ We used ADAM with an initial learning rate of $10^{-4},$ reducing it when reaching a plateau. We trained our network for 200 epochs with a batchsize of $5000.$

We have performed experiments with various architectures and different training objectives. We have varied the size of the hidden layers, the respective loss weights, and between multiplicative and additive metric corrections. We have also performed experiments on different complex structure ranges.

Unlike in the NNs with homogeneous coordinate inputs, we observe that the overlap is crucial. One possible explanation is the fact that we are training five independent networks which do not have to share any common property unlike in the homogeneous case where only one network deals with points from all patches. For these experiments we choose the points used to evaluate the overlap as follows. In order to make the patches the networks are defined on overlap, we slightly relax the numerical coordinate prescription. Instead of always dividing by the largest homogeneous coordinate such that all values are smaller than one, we allow values up to $1+\epsilon$. This guarantees that we do not require our neural network to make predictions very far away from where it is trained to make predictions on.

\bibliographystyle{bibstyle}
\bibliography{refs}
\end{document}